



\input harvmac.tex


\def\makeblankbox#1#2{\hbox{\lower\dp0\vbox{\hidehrule{#1}{#2}%
   \kern -#1
   \hbox to \wd0{\hidevrule{#1}{#2}%
      \raise\ht0\vbox to #1{}
      \lower\dp0\vtop to #1{}
      \hfil\hidevrule{#2}{#1}}%
   \kern-#1\hidehrule{#2}{#1}}}%
}%
\def\hidehrule#1#2{\kern-#1\hrule height#1 depth#2 \kern-#2}%
\def\hidevrule#1#2{\kern-#1{\dimen0=#1\advance\dimen0 by #2\vrule
    width\dimen0}\kern-#2}%
\def\openbox{\ht0=1.2mm \dp0=1.2mm \wd0=2.4mm  \raise 2.75pt
\makeblankbox {.25pt} {.25pt}  }

\def\bun#1/#2{\leavevmode
   \kern.1em \raise .5ex \hbox{\the\scriptfont0 #1}%
   \kern-.1em $/$%
   \kern-.15em \lower .25ex \hbox{\the\scriptfont0 #2}%
}

\def\opensquare{\ht0=3.4mm \dp0=3.4mm \wd0=6.8mm  \raise 2.7pt
\makeblankbox {.25pt} {.25pt}  }


\def\sector#1#2{\ {\scriptstyle #1}\hskip 1mm
\mathop{\opensquare}\limits_{\lower 1mm\hbox{$\scriptstyle#2$}}\hskip 1mm}

\def\tsector#1#2{\ {\scriptstyle #1}\hskip 1mm
\mathop{\opensquare}\limits_{\lower 1mm\hbox{$\scriptstyle#2$}}^\sim\hskip 1mm}


\def\bz{{\bar z}}
\def\ba{{\bar a}}
\def\bb{{\bar b}}
\def\bc{{\bar c}}
\def\bd{{\bar d}}

\def\ap{\alpha'}
\def\vol{{\rm vol}}
\def\IZ{{\bf Z}}
\def\IC{{\bf C}}
\def\IR{{\bf R}}
\def\mod{{\rm mod}}
\def\det{{\rm det}}
\def\p{\partial}
\def\pb{{\bar\partial}}
\def\CN{{\cal N}}

\def\zb{{\bar z}}
\def\tl{\tilde }
\def\imt{{{\rm Im}\tau~}}

\def\ap{A^{(+)} }
\def\am{A^{(-)} }
\def\op{\omega^{(+)} }
\def\om{\omega^{(-)} }
\def\CP{{\cal P}}

\def\a{\alpha}
\def\b{\beta}

\def\kab{K_{\a\b}}

\def\CG{{\cal G}}
\def\g{\gamma}

\def\L{{\Lambda}}
\def\tmab{\mu_{\a\b}}

\lref\AlvarezGaumeVM{
L.~Alvarez-Gaume, J.~B.~Bost, G.~W.~Moore, P.~Nelson and C.~Vafa,
``Bosonization On Higher Genus Riemann Surfaces,''
Commun.\ Math.\ Phys.\  {\bf 112}, 503 (1987).
}

\lref\bowditch{B.H. Bowditch, Geometrical finiteness for hyperbolic groups,
J. Funct. Anal. {\bf 113} (1993)245}

\lref\ManinHN{
Y.~I.~Manin and M.~Marcolli,
``Holography principle and arithmetic of algebraic curves,''
Adv.\ Theor.\ Math.\ Phys.\  {\bf 5}, 617 (2002)
[arXiv:hep-th/0201036].
}

\lref\WittenYA{
E.~Witten,
``SL(2,Z) action on three-dimensional conformal field theories with Abelian symmetry,''
arXiv:hep-th/0307041.
}

\lref\ElitzurNR{
S.~Elitzur, G.~W.~Moore, A.~Schwimmer and N.~Seiberg,
Nucl.\ Phys.\ B {\bf 326}, 108 (1989).
}

\lref\DeserWH{
S.~Deser, R.~Jackiw and S.~Templeton,
``Topologically Massive Gauge Theories,''
Annals Phys.\  {\bf 140}, 372 (1982)
[Erratum-ibid.\  {\bf 185}, 406.1988\ APNYA,281,409 (1988\ APNYA,281,409-449.2000)].
}
\lref\DeserVY{
S.~Deser, R.~Jackiw and S.~Templeton,
``Three-Dimensional Massive Gauge Theories,''
Phys.\ Rev.\ Lett.\  {\bf 48}, 975 (1982).
}

\lref\freedman{M. Freedman et. al. ``A class of P,T-invariant topological
phases of interacting electrons,'' cond-mat/0307511}

\lref\LarsenXM{
F.~Larsen,
``The perturbation spectrum of black holes in N = 8 supergravity,''
Nucl.\ Phys.\ B {\bf 536}, 258 (1998)
[arXiv:hep-th/9805208].
}

\lref\LarsenUK{
F.~Larsen and E.~J.~Martinec,
``U(1) charges and moduli in the D1-D5 system,''
JHEP {\bf 9906}, 019 (1999)
[arXiv:hep-th/9905064].
}

\lref\MaldacenaSS{
J.~M.~Maldacena, G.~W.~Moore and N.~Seiberg,
``D-brane charges in five-brane backgrounds,''
JHEP {\bf 0110}, 005 (2001)
[arXiv:hep-th/0108152].
}

\lref\CarlipBS{
S.~Carlip and I.~I.~Kogan,
``Three-Dimensional Topological Field Theories And Strings,''
Mod.\ Phys.\ Lett.\ A {\bf 6}, 171 (1991).
}

\lref\MooreYH{
G.~W.~Moore and N.~Seiberg,
 ``Taming The Conformal Zoo,''
Phys.\ Lett.\ B {\bf 220}, 422 (1989).
}
\lref\WittenHF{
E.~Witten,
``Quantum Field Theory And The Jones Polynomial,''
Commun.\ Math.\ Phys.\  {\bf 121}, 351 (1989).
}

\lref\gmms{
S.~Gukov, E.~Martinec, G.~Moore and A.~Strominger,
``The search for a holographic dual to AdS(3) x S(3) x S(3) x S(1),''
arXiv:hep-th/0403090.}

\lref\MooreVD{
G.~W.~Moore and N.~Seiberg,
``Lectures On Rcft,''
RU-89-32
{\it Presented at Trieste Spring School 1989}
}
\lref\AharonyTI{
O.~Aharony, S.~S.~Gubser, J.~M.~Maldacena, H.~Ooguri and Y.~Oz,
``Large N field theories, string theory and gravity,''
Phys.\ Rept.\  {\bf 323}, 183 (2000)
[arXiv:hep-th/9905111].
}

\lref\MooreQV{
G.~W.~Moore and N.~Seiberg,
``Classical And Quantum Conformal Field Theory,''
Commun.\ Math.\ Phys.\  {\bf 123}, 177 (1989).
}
\lref\MooreSS{
G.~W.~Moore and N.~Seiberg,
``Naturality In Conformal Field Theory,''
Nucl.\ Phys.\ B {\bf 313}, 16 (1989).
}
\lref\DijkgraafTF{
R.~Dijkgraaf and E.~Verlinde,
``Modular Invariance And The Fusion Algebra,''
Nucl.\ Phys.\ Proc.\ Suppl.\  {\bf 5B}, 87 (1988).
}
\lref\WittenWY{
E.~Witten,
``AdS/CFT correspondence and topological field theory,''
JHEP {\bf 9812}, 012 (1998)
[arXiv:hep-th/9812012].
}

\lref\zeereview{A. Zee, ``Quantum Hall Fluids,'' cond-mat/9501022}

\lref\moorewen{ Joel E. Moore, Xiao-Gang Wen,
``Classification of Disordered Phases of Quantum Hall Edge States,''
cond-mat/9710208 ;  Phys. Rev. B 57, 10138-10156 (1998)}

\lref\MaldacenaRF{
J.~Maldacena and L.~Maoz,
``Wormholes in AdS,''
arXiv:hep-th/0401024.
}

\lref\kutasovseiberg{D. Kutasov and N. Seiberg, ``More comments on string theory on
AdS(3),''  Published in JHEP 9904:008,1999
e-Print Archive: hep-th/9903219}
%

\lref\DiaconescuBM{
E.~Diaconescu, G.~Moore and D.~S.~Freed,
``The M-theory 3-form and E(8) gauge theory,''
arXiv:hep-th/0312069.
}

\lref\FuchsSQ{
J.~Fuchs, A.~N.~Schellekens and C.~Schweigert,
``Galois modular invariants of WZW models,''
Nucl.\ Phys.\ B {\bf 437}, 667 (1995)
[arXiv:hep-th/9410010].
}

\lref\GannonKI{
T.~Gannon,
``Boundary conformal field theory and fusion ring representations,''
Nucl.\ Phys.\ B {\bf 627}, 506 (2002)
[arXiv:hep-th/0106105].
}

\lref\FuchsCM{
J.~Fuchs, I.~Runkel and C.~Schweigert,
``TFT construction of RCFT correlators. I: Partition functions,''
Nucl.\ Phys.\ B {\bf 646}, 353 (2002)
[arXiv:hep-th/0204148].
}

\lref\read{N. Read, ``Excitation structure of the hierarchy scheme
in the fractional quantum Hall effect,'' Phys. Rev. Lett. {\bf 65}(1990)1502.}

\lref\nikulin{V.V. Nikulin, ``Integral Symmetric Forms and Some of Their Applications,''
Math. USSR Izvestiya {\bf Vol. 14}(1980) 103.}

\lref\DKSS{S.~Deger, A.~Kaya, E.~Sezgin and P.~Sundell,
``Spectrum of D = 6, N = 4b supergravity on AdS(3) x S(3),''
Nucl.\ Phys.\ B {\bf 536}, 110 (1998), hep-th/9804166.}

\lref\LPS{H.~Lu, C.~N.~Pope and E.~Sezgin,
``SU(2) reduction of six-dimensional (1,0) supergravity,''
hep-th/0212323.}

\lref\LPSb{H.~Lu, C.~N.~Pope and E.~Sezgin,
``Yang-Mills-Chern-Simons supergravity,'' hep-th/0305242.}

\lref\APT{G.~Arutyunov, A.~Pankiewicz and S.~Theisen,
``Cubic couplings in D = 6 N = 4b supergravity on AdS(3) x S(3),''
Phys.\ Rev.\ D {\bf 63}, 044024 (2001), hep-th/0007061.}

\lref\Mathur{S.~D.~Mathur,
``Gravity on AdS(3) and flat connections in the boundary CFT,''
hep-th/0101118.}

\lref\NS{H.~Nicolai and H.~Samtleben,
``Kaluza-Klein supergravity on AdS(3) x S(3),'' hep-th/0306202.}

\lref\ADPW{S.~Axelrod, S.~Della Pietra and E.~Witten,
``Geometric Quantization Of Chern-Simons Gauge Theory,''
J.\ Diff.\ Geom.\  {\bf 33}, 787 (1991).}

\lref\BNair{M.~Bos and V.~P.~Nair,
``U(1) Chern-Simons Theory And C = 1 Conformal Blocks,''
Phys.\ Lett.\ B {\bf 223} (1989) 61;
``Coherent State Quantization Of Chern-Simons Theory,''
Int.\ J.\ Mod.\ Phys.\ A {\bf 5} (1990) 959.}

\lref\Manoliu{M.~Manoliu,
``Abelian Chern-Simons theory,'' J.\ Math.\ Phys.\ {\bf 39} (1998) 170;
``Quantization of symplectic tori in a real polarization,''
dg-ga/9609012.}

%
%
\Title{\vbox{\baselineskip12pt
\hbox{hep-th/0403225}
\hbox{HUTP-04/A014}
}}
{\vbox{\centerline{
 Chern-Simons Gauge Theory   }
\centerline{and the $AdS_3/CFT_2$ Correspondence } }}
\centerline{Sergei Gukov\footnote{$^*$}{\it Jefferson Physical
Laboratory, Harvard University, Cambridge, MA 02138},
Emil Martinec\footnote{$^{**}$}{\it University of Chicago,
Chicago, IL 60637},
Gregory Moore\footnote{$^\dagger$}{\it Department of Physics and
Astronomy, Rutgers University, Piscataway, NJ 08855} and Andrew
Strominger$^*$} \vskip.1in \vskip.1in \centerline{\bf Abstract}
\noindent

The bulk partition function of pure Chern-Simons theory on a
three-manifold
 is a state in the space of
conformal blocks of the dual boundary RCFT, and therefore
transforms non-trivially under the boundary modular group. In
contrast the bulk partition function of $AdS_3$ string theory is
the modular-invariant partition function of the dual CFT on the
boundary. This is a puzzle because $AdS_3$ string theory formally
reduces to pure Chern-Simons  theory at long distances. We study
this  puzzle in the context of massive Chern-Simons theory. We
show that  the puzzle is resolved in this context  by the
appearance of a chiral ``spectator boson'' in the boundary CFT
which restores modular invariance. It couples to the conformal
metric but not to the gauge field on the boundary. Consequently,
we find a  generalization of the standard Chern-Simons/RCFT
correspondence involving ``nonholomorphic conformal blocks'' and
nonrational boundary CFTs. These generalizations appear in the
long-distance limit of $AdS_3$ string theory, where the role of
the spectator boson is played by other degrees of freedom in the
theory.

\Date{March 19, 2004}
\listtoc \writetoc

\def\IH{{\bf H}}
\def\S{{\bf S}}


\newsec{Introduction}

One of the most beautiful examples of a holographic correspondence
is the equivalence between three-dimensional Chern-Simons gauge theory
and the chiral half of a rational conformal field theory
 \WittenHF. (For reviews see \refs{\MooreVD,\ADPW,\Manoliu}).
We will refer to this as the CSW/RCFT correspondence.
In recent years a more ambitious example of holography has been
investigated, that of the AdS/CFT correspondence \AharonyTI. In this paper we
discuss some aspects of the relation between these two holographic
dualities.  

We expect to find a relation between
the AdS/CFT correspondence and the CSW/RCFT correspondence
in the special case of superstring theories on spacetimes of the form
$AdS_3 \times K_7$, where $K_7$ is a compact $7$-manifold. The reason is
that   the
low energy supergravity on $AdS_3$ typically contains gauge fields
with Chern-Simons terms. This raises
a puzzle when $K_7$ is a product of spheres,
such as $K_7 = \S^3 \times (\S^1)^4$ or
$K_7 = \S^3 \times \S^3 \times \S^1$,
because in those cases the dual conformal field theory
associated with the $\S^1$ factors
is in general {\it not} a rational conformal field theory. The
present paper resolves that puzzle.

In this paper we examine in some detail the holography of the massive
abelian gauge theories that appear in the $AdS/CFT$ examples we have
just cited. At long distances these theories are dominated by the
Chern-Simons terms. We will show that the partition function of these theories
has a kind of factorization into ``non-holomorphic conformal blocks,''
which transform in a finite-dimensional representation of the modular
group. They are associated to a theory of a {\it nonchiral} boson,
consisting of the usual chiral boson plus an antichiral ``spectator'',
and
have a continuously variable radius. We think this is an interesting extension of the
standard holographic duality of CSW theory to  the chiral half of a
rational conformal field theory.

Let us describe our results in some more detail. In section 2 we
review the well-studied example of a single massive abelian
gauge field  with  action \refs{\DeserWH,\DeserVY}
\eqn\stan{
S= \int {1\over 2e^2} dA*dA  -2\pi ik A dA.
}
(Here it is in euclidean signature; our
conventions are spelled out in the text below.)
The partition function  of the theory is a product of two factors; one
factor is  associated with
a massive scalar field, and the other with  a topological sector of the
theory. We are mainly interested in the latter, although we shall see that
the effects of the first term  do not entirely disappear
at long distances. The most natural way to study this
theory --- especially in the context of $AdS_3$ string theory --- is to compute
the path integral on a three-manifold $Y$ as a function of the boundary
conditions on the metric and gauge fields on $X:=\p Y$.


In this paper we focus on the quantization of the
theory on a  solid torus with Dirichlet boundary conditions for the gauge fields.
We consider the limit of an infinite-volume torus (such as a quotient by $\IZ$ of
hyperbolic space). In this limit
we can study the partition function by studying the
groundstate of the gauge theory on $T^2$. We do this by solving explicitly
for the Landau level wavefunctions in the quantization on the plane, and then
projecting onto gauge invariant wavefunctions, taking proper account of the
Gauss law. In this way we produce a finite dimensional space of wavefunctions,
and the partition function on the torus is a linear combination of these
wavefunctions.

In the above approach
it turns out to be important to include both chiralities of the boson on
the boundary, although only one of these couples to the gauge field -this being
the usual chiral boson of CSW theory.
Put differently, the partition function on the solid torus, in the limit
of infinite volume, is equivalent to that of a nonchiral boson with
Euclidean action:
\eqn\gaussmd{
\pi k  \int d\phi * d \phi+ 4\pi i k \int \pb \phi \wedge A^{1,0}
}
where $\phi \sim \phi + 1 $ and therefore the target space of the
boson is a circle of radius $R^2 = k \alpha'$.
(We have assumed $k>0$). The fact that we can even speak of the
radius shows that we must include both left- and right-movers.
We refer to the left-moving part of $\phi$, which is invisible to $A$,
as a ``spectator chiral boson.'' Note that the spectator does
couple to the conformal metric on the boundary.

In section 3 we turn to the main model of interest here, namely the
theory of two abelian gauge fields with off-diagonal
Chern-Simons coupling. The action is:
\eqn\abthry{
S_a =
\int {1\over 2e_A^2} dA*dA + {1\over 2e_B^2} dB*dB -2\pi ik A dB
}
Our primary motivation is that this is the form of Lagrangian appearing in
the low-energy supergravity theory on $AdS_3$ in the examples cited above,
although as we discuss at the end of this introduction, there are other
potential applications of our remarks.

The topological sector of the theory has two parameters, these are the integer
$k$ and the real number $\mu := \vert e_B/e_A\vert$.
One might think that \abthry\ is a trivial extension of \stan\ since
one could introduce the change of variables
\eqn\invcoms{
\eqalign{
A & = {1\over \sqrt{2\mu}}(\ap - \am) \cr
B & = \sqrt{\mu\over 2} (\ap + \am) \cr}
}
which gives two copies of \stan, but with $e^2 \to \vert e_A e_B\vert $, and
 $k \to +\half k$ for
one term while  $k \to - \half k$ for the other.
It turns out that we do not get a trivial extension of the previous
theory, because of the quantization conditions on the periods of $A$ and
$B$.
The dual theory is a
theory of two bosons, $\phi^A, \phi^B$ of period $1$ with
Euclidean action  of the form $S_1 + S_2 + S_3$ where
\eqn\quadrt{
S_1 = {\pi k \over 2} \int \mu d\phi^A * d \phi^A + \mu^{-1} d\phi^B * d \phi^B
}
shows the bosons have radius $R_A^2 = \half k \mu \alpha'$ and $R_B^2 = \half k \mu^{-1} \alpha'$
while
\eqn\bfield{
S_2 =  i \pi k \int d\phi^A \wedge d \phi^B
}
shows there is a nontrivial $B$-field, for $k$ odd, and finally
\eqn\linact{
S_3 = 2 \pi i k   \int \bigl[ (A^{(-)})^{0,1} \wedge \p \phi^{(-)} - (A^{+})^{1,0} \wedge \pb \phi^{(+)}\bigr]
}
gives the coupling to the gauge fields. In conformity with \invcoms\ we have defined
\eqn\linecombsi{
\eqalign{
\phi^{(+)} & :={1\over \sqrt{2}}( \mu^{-1/2} \phi^B +  \mu^{1/2}  \phi^A)\cr
\phi^{(-)} & := {1\over \sqrt{2}}( \mu^{-1/2} \phi^B -  \mu^{1/2}  \phi^A) \cr}
}
Note that $\phi^{(-)}_L + \phi^{(+)}_R$ is a nonchiral scalar coupling to the
gauge fields, but, for $\mu$ non-rational, it does not have a well-defined
periodicity, as promised.

The radii satisfy\foot{In what follows, $\alpha'=2$ unless noted otherwise.}
\eqn\rarb{
\eqalign{
{R_A \over R_B} & = \mu \cr
 R_A R_B & = k \cr}
}
Although the   boundary conformal field theory is {\it not} rational (when
$\mu$ is not rational), thanks to the quantization of $R_A R_B$,
  the partition function on the torus
is always a linear combination:
\eqn\zeelin{
Z = \sum_{\beta \in \Lambda^*/\Lambda}  \zeta^\beta \Psi_\beta(A, B) .
}
Here $\Psi_\beta(A,B)$ span a finite-dimensional space of states.
They are proportional to Siegel-Narain theta functions (defined in 
appendix A) associated to the
hyperbolic lattice $\Lambda= \sqrt{k} II^{1,1}$, and
\eqn\betas{
\beta \in \Lambda^*/\Lambda \cong (\IZ/k\IZ)^2 .
}
The $\Psi_\beta$  are also not holomorphic in $\tau$, but do transform in a simple
finite dimensional representation of the modular group. These higher level
theta functions generalize the familiar
holomorphic level $k$ theta functions of RCFT.
The case $k=1$ is simply   the modular invariant partition function of
a single compact boson.

In section 4 we show that our
 considerations easily extend to the  most general abelian Chern-Simons
theory with gauge group $U(1)^d$ and action
\eqn\manyflds{
\int{1\over 2 e^2}  \lambda^{-1}_{\a\b} dA^\a * dA^\b - 2\pi i K_{\a\b}   A^\alpha dA^\beta
}
where $2K_{\a\b} $ is an even integral nondegenerate symmetric matrix,
and hence defines a lattice $\bar \L$, while $\lambda^{\a\b}$ is a positive
definite symmetric matrix. The boundary theory, including the spectator chiralities,
is a theory of $d$ nonchiral bosons. The metric for the bosons is determined by
$\lambda^{\a\b}$ and $\kab$ while  the B-field is
\eqn\manybee{
- 2\pi i\int \sum_{\a<\b} K_{\a\b} d\phi^\a \wedge d \phi^\b
}
Left plus  right movers move in a target space
$V_L \oplus V_R$, where $V\cong \IR^d$. Using the data of
both  $\lambda^{\a\b}$ and $\kab$
one constructs a projection matrix $P_\pm$ on $V$ which is compatible with the
projection into left and right movers. The bosons coupling to the
gauge fields lie in $V_{L,-} \oplus V_{R,+}$. The ``spectator chiralities'' lie in
$V_{L,+}\oplus V_{R,-}$.

Finally, the computations also generalize in a natural way
from the torus to higher genus Riemann surfaces.

Now let us discuss the relation to the purely topological
CSW theory. In the $AB$ theory,
the  space of states spanned by $\Psi_\beta$ in \zeelin\ is $k^2$-dimensional, in
harmony with a standard analysis of the pure Chern-Simons theory associated to
the $e_A^2, e_B^2 \to \infty$ limit of \abthry\
\refs{\WittenHF,\BNair,\ElitzurNR,\WittenYA}.
Indeed, the  topological Hilbert space and the representation of the modular group
are independent of $\mu$ (and independent of $\lambda^{\a\b}$ in the higher
rank case). Nevertheless, the path integral on the torus naturally
introduces $\mu$-dependence in the basis of wavefunctions,
and is essential in writing the path integral of the massive Chern-Simons theory.
The dependence of the topological field theory on $\mu$ is quite analogous to
the dependence of the topological Hilbert spaces $\CH(X)$ associated to a
Riemann surface $X$ on the complex structure of $X$.
Because it is the fields   $\ap_z, \am_\zb$ which couple to the currents,
the holomorphic polarization is more natural when using the AdS/CFT correspondence.
Indeed, the path integral on $AdS_3$ with no operator insertions is in the
state \zeelin\ with $\zeta^\beta\sim \delta_{\beta,0}$.

Our work touches on a number of other closely related investigations.
First it touches on an old problem in the CSW/RCFT correspondence.
The chiral half of an RCFT is only part of the data needed to
construct the conformal field theory, as stressed
in \refs{\MooreQV,\MooreSS}. Indeed,
in general different CFT's can be made from gluing together
the chiral parts using different automorphisms of the fusion rules
\refs{\MooreSS,\DijkgraafTF}.
\foot{There are important subtleties in this statement which have
been investigated in \refs{\FuchsSQ,\GannonKI,\FuchsCM}. However they do not
affect the very simple models considered in this paper. }
Thus, a vexing question  has always been:
``How does one modify the CSW theory to incorporate both left- and
right-movers?'' The present paper provides the beginning of an
answer to that question, at least  in the case of abelian gauge theories.

The importance of including the kinetic terms $\sim \int F*F $
 in studying the holography  of abelian Chern-Simons theory
was stressed by S. Carlip and Y. Kogan  in their attempt to
rewrite string theory as a topological membrane theory \CarlipBS.  They did
not explain the role of left- and right-movers in the way we are
doing, but introduced this term to account for dependence on the
boundary conformal structure. More recently, off-diagonal Chern-Simons
terms have been discussed by Witten in \WittenYA. In his discussion
it is crucial that the theory with $k=1$ is ``trivial.'' What this
means, in our context, is that there is only one wavefunction $\Psi_\beta$,
and it transforms trivially under the modular group. Indeed,
as we have noted, the level $1$ Siegel-Narain theta function is simply the
theta function appearing in  the modular invariant partition
function of a conformal field theory of
both left- and right-movers.

\lref\FrohlichVQ{
J.~Frohlich {\it et al.},
``The Fractional Quantum Hall Effect, Chern-Simons Theory, And Integral
Lattices,''
ETH-TH-94-18
}
\lref\FrohlichAK{
J.~Frohlich, U.~M.~Studer and E.~Thiran,
``A Classification of quantum Hall fluids,''
KUL-TF-94-35
}
\lref\FrohlichQM{
J.~Frohlich, T.~Kerler, U.~M.~Studer and E.~Thiran,
``Structure the set of imcompressible quantum hall fluids,''
Nucl.\ Phys.\ B {\bf 453}, 670 (1995).
}
\lref\BlokAN{
B.~Blok and X.~G.~Wen,
``Effective Theories Of Fractional Quantum Hall Effect: The Hierarchy
Construction,''
Phys.\ Rev.\ B {\bf 42}, 8145 (1990).
}

The present computations might conceivably find a use in condensed
matter physics, where classification of quantum hall states involves
the study of general abelian Chern-Simons gauge theories
\refs{\read, \BlokAN,\FrohlichAK,\FrohlichQM,\FrohlichVQ, \zeereview,\moorewen}.
Curiously, for related reasons, massive Chern-Simons theories with
two gauge fields and opposite sign Chern-Simons terms
have recently been recognized as being important in
condensed matter physics with a view towards quantum computation.
See, for example, \freedman.\foot{ We thank Paul Fendley for pointing this
out. }
Also in \WittenYA\ Witten pointed out that the triviality of the AB theory for
$k=1$ has important consequences for the classification of quantum Hall
states. In the simple case where we do not consider
spin theories, the   results of this paper, combined with the Nikulin
embedding theorem \nikulin\ show that abelian Chern-Simons theories
 are classified by the  signature of $\Lambda$ modulo 24, together with  
the discriminant form of the lattice $\Lambda$, where $\Lambda$ is the
lattice  determined by $-2\kab$.

Finally, one motivation for the present work was a
project involving the $AdS/CFT$ correspondence, so let us
mention  briefly here some implications for the $AdS/CFT$
correspondence. (Further details   are in \gmms.) The
relevance of topological field theories to the AdS/CFT
correspondence was first discussed in \WittenWY.
The authors of \MaldacenaSS\ discussed in some detail the
singleton sector of supergravity theories in the AdS/CFT
correspondence in a variety of dimensions. 
We will improve on \MaldacenaSS\ in two ways. 
First, we show that the Hamiltonian
for the singleton is naturally chosen by using Dirichlet 
boundary conditions for the second order system  in the 
Euclidean path integral. Second we show how one can discuss  
 the radius of the singleton scalar. 

We have studied here the simple free-field theory of abelian gauge fields.
In the AdS/CFT context these gauge fields   couple to other degrees of
freedom in the low energy supergravity. Nevertheless, based on simple considerations
of the decoupling of topological modes at long distance, we conjecture
that the full partition
function of the string theory on $AdS_3 \times K_7$ can still be written as
\eqn\zeestring{
Z_{\rm string} = \sum_\beta \zeta_{\rm string}^\beta \Psi_\beta(A,B)
}
%
where $\zeta_{\rm string}^\beta $ is $A,B$-independent, and $\Psi_\beta(A,B)$
are the same functions as in \zeelin. 
That is, the dependence on the boundary values of the $U(1)$ gauge fields is
given by a wavefunction in the topological Hilbert space determined by
the free massive gauge theory.  The essential difference from the massive
gauge theory (which is {\it not} a holographic theory, since it does not contain
gravity), is that $\zeta^\beta_{\rm string}$ only depends on boundary data.
In \gmms\ the
conjecture \zeestring\ is used to investigate the holographic
correspondence for $AdS_3 \times \S^3 \times \S^3 \times \S^1$.

\newsec{Review of the standard case}

\lref\SchwarzAE{
A.~S.~Schwarz,
Commun.\ Math.\ Phys.\  {\bf 67}, 1 (1979).
}
%

The massive 3d gauge theory was analyzed in a classic paper \refs{\DeserWH,\DeserVY}
and the topological sector of the theory was understood
in \refs{\SchwarzAE,\WittenHF,\MooreYH,\BNair,\ElitzurNR}.
We review it here as preparation for  the $AB$ theory.

\subsec{The classical theory}

We are interested in studying abelian Chern-Simons gauge theory
on a topologically non-trivial 3-manifold $Y$.
In this section, we focus on the simplest   example of such a
theory, with $U(1)$ gauge group, whose action is
(in   Euclidean signature),
\eqn\stane{
S_E= \int {1\over 2e^2} dA*dA - 2\pi i k A dA
}
Here, the gauge connection $A$ is a section of a principal
$U(1)$-bundle over $Y$, normalized so that $dA$ has integral
periods and large gauge transformations are $A \to A+ \omega$
with $\omega$ a closed 1-form with integral periods.
In order to obtain the partition function of the Euclidean
theory, one has to integrate over the space of all gauge
connections $A$ (modulo the gauge equivalence) with the measure $e^{-S}$.
Similarly, on the Lorentzian space-time with signature $(-,+,+)$
the action looks like
\eqn\stan{
S= \int {-1\over 2e^2} dA*dA + 2\pi k A dA
}
and the measure is given by $e^{i S}$.

The coupling $k$ is an integer
if the Euclidean theory is to be well-defined on all 3-manifolds.
%
%
If we use spin bounding 4-folds
then
we can take $k_{min}=\pm \half$.

The coupling $e^2$ has dimensions of mass. Under a conformal rescaling
$g_{\mu\nu} \to \Omega^2 g_{\mu\nu}$ of the 3-dimensional metric the
first term in the action scales as $\Omega^{-1}$, while the second is
invariant. Therefore, we expect that at long distances the topological
term dominates. Note that in this sense the long-distance limit is the
$e^2 \to \infty$ limit.

The equations of motion are
\eqn\eom{
d*F - 4\pi k e^2 F =  0
}
In the presence of a boundary
we vary in a space of fields such that the two form 
\eqn\boundary{
 \delta A \wedge (* F - 2\pi k e^2 A) =0
}
vanishes
when pulled back  to the boundary.
 
\subsec{ Solutions to equations of motion}

We are interested in formulating carefully the phase space of the theory.
One way of formulating physical phase space is that it is the space of
gauge inequivalent solutions of the equations of motion.

In the present theory, thanks to linearity the space of (not necessarily
gauge inequivalent) solutions of the equations of motion is a product
\eqn\rpd{
S = S_f \times S_{nf}
}
where $S_f$ is the space of flat solutions $F=0$.   These are the solutions of the
topological sector. $A= A_f + A_{nf}$ where $A_{nf}$ is orthogonal to the flat
subspace in, say, the Hodge metric.

More generally, on $Y = X \times \IR$, $X$ compact we can take $ A= A_f + A_{nf}$
where the nonflat component $A_{nf}$ is defined by saying it is orthogonal to
$\ker d$ in the Hodge metric. The space of solutions to the equations of motion is
a {\it product}. When $X$ is noncompact one needs to include boundary
conditions, and the space of solutions might or might not be a product.

The main result of \refs{\DeserWH,\DeserVY} is the ``equivalence'' of the
massive gauge theory to a theory of a massive scalar field. In our
context this means that we can identify the factor $S_{nf}$ with the
space of solutions of the massive scalar equation.

\subsec{ Hamiltonian Formalism}

Let us work out the Hamiltonian formulation   on a spacetime
of the form   $X \times \IR$,
with metric $-dt^2 + g_{ij} dx^i dx^j$ and orientation
$dt dx^1 dx^2$.  The canonically conjugate
momentum as a vector-density is ($\epsilon^{12}=+1$):
\eqn\ccmom{
\Pi^i = {1\over e^2} \sqrt{g} g^{ij}(\dot A_j -\p_j A_0) + 2\pi k \epsilon^{ij} A_j
}
The action can be written as $S= \int dt L $ with
\eqn\writeact{
L =   \int_X \Pi^i \dot A_i -     H  +   \int_X A_0
\biggl( \p_i \Pi^i + 2\pi k \epsilon^{ij} \p_i A_j \biggr)
}
where we find  a Hamiltonian  
\eqn\hamdens{
H = \int_X {e^2\over 2 \sqrt{g}} g_{ij} E^i E^j + {1\over   2e^2} F\wedge *_2 F
}
where  $*_2$ is the Hodge star on $X$ and
\eqn\tildps{
E^i := \Pi^i - 2\pi k \epsilon^{ij}A_j
}
(We will also denote $E^i = \tilde \Pi^i$ below.)
The Gauss law is:
\eqn\gaussclass{
\p_i \Pi^i + 2\pi k \epsilon^{ij} \p_i A_j =0
}
That is, $\nabla \cdot E + 4\pi k B =0 $.

\subsec{Phase space and symplectic structure}

There are two descriptions of the phase space, depending on
how one works with Hamiltonian reduction.

One way to formulate physical  phase space is as the space of gauge inequivalent solutions of
the equations of motion. This point of view makes it obvious that the phase space is
a product of the phase space for flat gauge fields and for nonflat gauge fields,
 $\CP = \CP_f \times \CP_{nf}$ for the flat and nonflat
parts of the theory.

Another  way to formulate the theory ``upstairs'' in $A_0=0$ gauge is to
take phase space to be the cotangent space with coordinates
 $(\Pi^i, A_i)$ and symplectic form:
\eqn\omebig{
\Omega  = \int_X \delta \Pi^i \wedge \delta A_i
}
where $\delta $ is exterior derivative on the infinite dimensional phase space.
Notice that when \omebig\ is restricted to the subspace of flat gauge fields, by
\tildps\  we get second class constraints and the phase space is
the Chern-Simons symplectic form
\eqn\omeflat{
\Omega_f  = \int_X  2\pi k \delta A  \wedge \delta A
}
This is gauge invariant on the subspace $F=0$ and one may then perform Hamiltonian
reduction.

 It is instructive to consider the   $e^2\to \infty $ limit.
Using \hamdens,  we see that if we restrict
to finite energy field configurations then we must set $E^i =0$. Then, by the
Gauss law we must put   $F=0$. As we have said, restriction to this subspace
imposes second class constraints  and we are restricting to
the flat factor in phase space.

\subsec{Quantization in the Schrodinger representation}

If we quantize on phase space and then impose the Gauss law we have
wavefunctionals $\Psi(A_i) $, and we quantize using the symplectic
form \omebig.  Thus
\eqn\schr{
\Pi^i = - i {\delta \over \delta A_i}
}

Since we can split $A= A_f + A_{nf}$ and the Hamiltonian does not mix
these, the Hilbert space of the
theory is naturally thought of as a product
\eqn\fctrz{
\CH =  \CH_f \otimes \CH_{nf}
}
where $\CH_f$ is the space of wavefunctions of flat potentials.

The  Gauss law is:
\eqn\gauslaw{
\Psi(A + \omega) = e^{-2\pi i k \int \omega \wedge A } \Psi(A)
}
This is valid also for large gauge transformations.
\foot{This requires explanation. The proper mathematical formulation
involves regarding $\Psi$ as a section of a line bundle over the space of
gauge potentials $\CA(X)$ on $X$. We then lift the group action, and find
that a lift only exists when $c_1(P)=0$. There is a canonical trivialization
of the line in this case, as well as a canonical connection, and the
wavefunction becomes a function. A similar discussion holds for the more
subtle case of the M-theory C-field \DiaconescuBM.}
Here $\omega$ is a closed 1-form with integral periods. Note that this
does not affect the $A_{nf}$ variable.

\subsec{Euclidean Path integral on the solid torus }

We will determine the Hamiltonian for the singletons by considering
the Euclidean path integral
of the theory on the solid torus, and then interpreting that path integral in terms
of Hamiltonian evolution in the radial direction.

Since our action is second order in derivatives, when formulating
the path integral on a handlebody $Y$ we should
specify   all components of $A_X$ on the
boundary $X$.  This is to be contrasted with the Chern-Simons path integral
which is a phase space path integral, and in which we specify just
  one component of $A_X$ on the boundary $X$.

Let us consider  the Euclidean partition function of the theory
on a solid torus with radius $\rho$ denoted $Y_{\rho} \cong D \times \S^1$.
We assume the torus has a metric
that behaves asymptotically like $d\rho^2 + \Omega^2(\rho) g_X $. The path integral
defines a state $\Psi_{Y_\rho}(A)$ given by
\eqn\pathstate{
\Psi_{Y_\rho}(A) = \int  {dA_Y \over \vol(\CG(Y))} e^{-\int {1\over 2e^2} dA*dA + 2\pi i k\int  A dA }
}
where $\CG(Y)$ is the gauge group on $Y_{\rho}$. We can  understand the behavior
for $\rho\to \infty$ just from the above understanding of the spectrum.

We can view the evolution to large $\rho$ as evolution in a Euclidean
time direction. The large $\rho$ behavior projects onto the lowest
energy states.
\eqn\projlll{
\lim_{\rho\to \infty} \Psi_{Y_\rho}(A) = e^{-\rho E_0} \Psi_0
}
with $\Psi_0$ in the space of ground states on the torus with energy $E_0$. The insertion of
local operators such as  Wilson lines or other disturbances induces transitions between
vectors within this space of ground states.

\subsec{Quantization on the torus}

We now consider quantization on $T^2 \times \IR$.
Our wavefunction is $\Psi(A_f) \otimes \Psi(A_{nf})$. The spectrum of the
nonflat sector is clear, and we take the unique groundstate wavefunction
for this factor: It is the product of harmonic oscillator groundstates for
the oscillators of the massive scalar of \refs{\DeserWH,\DeserVY}. In this
section we drop this factor
 so we can focus on the dependence on $A_f$.

To simplify matters, we work on a torus $X= T^2$ with $z = \sigma^1 + \tau \sigma^2$
and metric $\Omega^2 \vert dz \vert^2$, $\sigma^i \sim \sigma^i + 1$.
We fix the small gauge transformations by assuming $A_f$ is constant.

In complex coordinates $A = A_z dz + A_\zb d\zb$ we have
$$
A_z = {A_2 - \bar \tau A_1 \over \tau-\bar \tau} \qquad A_\zb = -
{A_2 - \tau A_1 \over \tau-\bar \tau}.
$$
We further define the zero mode of the shifted momentum \tildps\
as
\eqn\compsl{ \eqalign{ \tilde \Pi^z & =\int d^2z\bigl(-i{\delta
\over \delta A_z(z)}-2\pi k\epsilon^{z \bar z}A_{\bar z}(\bar
z)\bigr)= - i \left( {\p \over \p A_z} - 4\pi k \imt A_{\bar z}
\right) \cr \tilde \Pi^\zb & = - i \left( {\p \over \p A_\zb} +
4\pi k \imt A_{ z} \right) \cr } }
so that the Hamiltonian   is:
\eqn\hamdensi{
H = {e^2\over 4\imt} (\tl \Pi^z \tl \Pi^\zb + \tl \Pi^\zb \tl \Pi^z)
}
Note that these do not commute: $[\tilde \Pi^z, \tilde \Pi^\zb] = - 8 \pi k \imt $.
The ground state energy density is $2\pi \vert k \vert e^2$ and is infinitely
degenerate, as in the standard Landau-level problem.
If $k>0$ we have
\eqn\psoa{
\tl \Pi^\zb \Psi=0  \Rightarrow \Psi = e^{-4\pi k \imt A_z A_\zb} \psi(A_z)
}
If $k<0$ we have
\eqn\psob{
\tl \Pi^z \Psi=0  \Rightarrow \Psi = e^{4\pi k \imt A_z A_\zb} \psi(A_\zb)
}
Here $\psi$ are arbitrary holomorphic functions. Indeed, if we take $\psi= \psi_\lambda$, where
\eqn\coherent{
\psi_\lambda(x) := e^{\lambda x}
}
then the set of   wavefunctions $\{\Psi_\lambda\vert \lambda\in \IC\}$
 is an overcomplete
set spanning the infinite-dimensional lowest Landau level.

The set of states spanned by \coherent\ is infinite dimensional, but when we
consider gauge invariant wavefunctions on the torus
the lowest Landau level (LLL)
becomes finite
dimensional.  We have already enforced the
invariance under small gauge transformations by choosing our flat connections
to be constants on the torus. We can impose the invariance under large gauge
transformations by averaging over large gauge transformations.
  Given {\it any} wavefunction $\Psi(A)$
the average:

\eqn\average{
\bar \Psi(A) := \sum_{\omega \in {{\rm Harm}}^1_Z} \Psi(A+\omega) e^{2\pi i k \int \omega A}
}
where ${{\rm Harm}}^1_Z$ are the harmonic 1-forms with integral periods,
transforms according to the Gauss law \gauslaw.

Now assume $k>0$ and  get the projected wavefunctions of the LLL:
\eqn\averagei{
\bar \Psi(A) := \CN e^{-4\pi k \imt A_z A_\zb}
\sum_{\omega \in {{\rm Harm}}^1_Z}
e^{-4 \pi k \imt \omega_z \omega_\zb} e^{-8 \pi k \imt \omega_\zb A_z}
\psi(A_z + \omega_z)
}
where
$\CN$ is a normalization constant, which might depend on $\tau$.

Let us now consider the space of wavefunctions --- as functions of $A_z$ --- that
 we obtain from \averagei\ and
\coherent. At first, one might think that the space is infinite dimensional
since $\lambda$ in \coherent\ can be any complex number.
However, using the Poisson summation formula we find that
\eqn\psftmrn{
\overline{\Psi_\lambda} = e^{-4\pi k \imt (A_z A_\zb + A_z^2) - {\lambda^2\over 16 \pi k \imt}  }
\sqrt{\imt \over k}
 \sum q^{\half p_L^2 } \bar q^{\half p_R^2} e^{- {p_R \over R} 8\pi k \imt A_z - {p_L \over R} \lambda}
}
where $q=e^{2\pi i \tau}$, and
\eqn\dfns{
\eqalign{
& p_L = (n/R + mR/2) , p_R = (n/R - mR/2) \cr
& R^2 = 2k \cr }
}
We recognize that we have the soliton sum of the  partition function of a scalar field
with radius $R$. Since $R^2 = 2k$ is integral it is a rational conformal field
theory, and the infinite sum can be split as a finite sum of terms of
the form $f_i(A) g_i(\lambda)$.
The sublattices $p_L=0$ and $p_R=0$ are of index $2k$ in the Narain lattice
$(p_L; p_R)$. Indeed, after a little algebra we see that \psftmrn\ can be written
as:
\eqn\psftmrni{
\overline{\Psi_\lambda }= e^{-Q} 
\sqrt{\imt \over k} \sum_{0 \leq \mu < 2k }
\Theta_{-\mu,k}(-2i \imt A_z,-\bar\tau)  \Theta_{\mu,k}({-\lambda\over 4\pi i k}, \tau).
}
where 
\eqn\extrq{
Q = 4\pi k \imt (A_z A_\zb + A_z^2) + {\lambda^2\over 16 \pi k \imt} 
}
The level $k$  theta functions $\mu=-k+1, \dots, k$ are defined by
\eqn\xxz{\Theta_{\mu,k}(\omega,\tau)=\sum_{n\in \IZ} q^{k(n+\mu/(2k))^2}
y^{ (\mu + 2k n)}}
where $y = \exp(2\pi i \omega)$. 
Equation \psftmrni\ shows quite explicitly that the space of quantum states is in fact
only finite dimensional. A basis for the vector space of states is
\eqn\basis{
\psi_\mu = \CN \sqrt{\imt \over k}  e^{-4\pi k \imt (A_z A_\zb + A_z^2) } \Theta_{-\mu,k}(-2i\imt A_z,
-\bar \tau) \qquad 1\leq \mu\leq 2k
}
Finally, we would like to determine the normalization $\CN$. We do this following a
trick in \ElitzurNR.

The flat gauge fields on the torus can be written $A = d \chi + A_z dz + A_\zb d \zb $.
{}From the Gauss law $\Psi(A) = \psi(A_z, A_{\zb})$. However there is a Jacobian for
the change of variables from $A$ to $\chi, A_z, A_\zb$.
Now
\eqn\innpr{
\int {[dA] \over \vol \CG} \Psi_\mu^*[A] \Psi_\nu[A] = \det'(d) \int_0^1 dA_1 \int_0^1 dA_2
\psi_\mu^* \psi_\nu
}
We can regularize $\det'(d) = \sqrt{\imt} \vert \eta\vert^2$. We can also evaluate the inner product
of the states \basis:
\eqn\ipbsis{
\eqalign{
\int_0^1 dA_1 \int_0^1 dA_2  \psi_\mu^* \psi_\nu & = \delta_{\mu,\nu}
{\imt \over k} \vert \CN \vert^2
\int_0^1 dA_1 \sum_n e^{-4\pi k \imt (A_1 + n- \mu/2k)^2} \cr
& = {\sqrt{\imt} \over 2k^{3/2}} \vert \CN \vert^2\cr}
}
Normalizing the wavefunctions to one gives:
\eqn\basis{
\psi_\mu = {k^{3/4}\over \bar \eta}   e^{-4\pi k \imt (A_z A_\zb + A_z^2) } \Theta_{-\mu,k}(-2i\imt A_z,
-\bar \tau)
}

{\bf Remarks:}

\item{1.} The higher Landau levels are obtained by acting with
$\tl \Pi^z$ to give energy densities $2 \pi k e^2 (2N+1)$, $N>0$. Note that
\averagei\ is independent of $e^2$, and hence has a smooth limit
as $e^2\to\infty$. Moreover, the gap between Landau levels becomes infinite
in this limit.

\item{2.}
The dependence on $A_z$ is that of the wavefunctions in
the holomorphic polarization of the pure Chern-Simons theory.
Equivalently, they are conformal blocks for the Gaussian model
at $R^2=2k$, coupled to an external gauge field.

\subsec{ Holographic mapping to the Gaussian model}

We now interpret the sum \averagei\ in terms of conformal field
theory. The first
exponential factor in the sum in
\averagei\   is just the standard value of the Gaussian model
 action
\eqn\newgaussmd{
k \pi  \int d\phi * d \phi
}
evaluated on a soliton configuration $d\phi = \omega = n_1 d\sigma^1 + n_2 d \sigma^2$.
In our conventions, we use a scalar of periodicity $1$ and hence we get the
Gaussian model on a circle with radius
\eqn\effctv{
R^2 = k \alpha',
}
{\it with both left-movers and right-movers}. Of course, we then recognize the
Narain lattice in \dfns\ with $\alpha'=2$.
Note, however, that the coupling of $\phi$ to $A_z$ is chiral, and given by
the Lagrangian
\eqn\coupling{
4\pi i k \int \pb \phi \wedge A^{1,0}
}

For $k>0$ we have holomorphic functions of $A_z$ coupling to the rightmoving
current $\bar \p \phi$ and for $k<0$ we have holomorphic functions of $A_\zb$
coupling to the leftmoving current $\p \phi$.

{\bf Remarks: }

\item{1.}
  In   \CarlipBS\ Carlip and Kogan discuss very closely related matters in their
attempt to rewrite string theory as a topological membrane theory.  The Landau
levels are solved for in their eq. 3.4, which they are thinking of as
the solutions of the
 full Schrodinger equation in the limit $e^2\to \infty$. Their motivation
was to introduce dependence on the conformal structure of the boundary into
the wavefunctions. They intended to get left and right-moving degrees of
freedom from the inner and outer radii of an annulus,
as in \refs{\MooreYH,\ElitzurNR}.

\item{2.} We now propose a somewhat heterodox interpretation of the
equation \averagei. We propose that the dual conformal field theory
is a theory of both a left-moving and a right-moving boson with the
radius \effctv. The fact that both chiralities are present is surprising
since the canonical quantization of the pure Chern-Simons theory is well-known
to lead to a single chiral boson. In particular, with 
appropriate boundary conditions the quantization
on the disk gives a chiral boson degree of freedom on the boundary. One should
distinguish between the {\it modes of $A$ on the boundary} which, with proper boundary
conditions are those of a chiral scalar and {\it the dual field theory variable $\phi$}.
Note that $\p \phi$ {\it couples} to $A$, it is not one of the degrees of freedom
of $A$. Moreover, only
 one chirality of $\phi$ couples to $A$. The other chirality  is a ``spectator'' in the
sense that in \averagei\  only one chirality $\omega_\zb$ couples to the external
gauge field. Nevertheless, there are really two chiralities present
in \averagei. Both chiralities couple to the conformal metric.

\item{3.} One way to make the point about the ``reality'' of the spectator
chirality is to note that   we have identified a definite radius, \effctv.
In order to understand why this is surprising one must recall some standard
points from RCFT.
In RCFT the wavefunctions \basis\ are the conformal blocks of a Gaussian model
with ``$U(1)$ level $N$ current algebra.'' By definition, this is a holomorphic $U(1)$
current algebra extended by holomorphic currents
\eqn\uoneext{
e^{\pm i \sqrt{2N} \phi(z)}
}
of conformal dimension $N$. There are $2N$ distinct representations of this
algebra generated by $\exp[ i {r\over \sqrt{2N}} \phi(z)]$ for $r \sim r+ 2N$.
The conformal blocks of this theory on the torus
are level $N$ theta functions. This theory is dual to the pure Chern-Simons
theory with action
\eqn\uones{
2 \pi i N \int A d A
}
in a normalization where $dA$ has integer periods. In units $\alpha'=2$ the
Gaussian model with radius $R$ has $U(1)$ level $N$ current algebra whenever
$R^2$ is rational. More precisely, if $R^2=p/q$ is in lowest terms then
$N = 2pq$ for $p$ odd and $N=pq/2$ for $p$ even. In the present section
we have $R^2=2k$ and hence $N=k$, hence $2k$ topological states.
Returning to the general case,  for a given $N$ there are
several Gaussian models with the same chiral algebra, corresponding to the
different ways of factoring $N$. The choice of a definite radius is a
choice of how to combine  left- and right-moving conformal blocks
\refs{\MooreSS,\DijkgraafTF}. One cannot speak of the
radius without introducing both left and right movers.

\item{4.} The role of the spectator chiralities is further clarified if
one compares carefully the Euclidean and Lorentzian versions of holography.
In the Lorentzian case we have an isomorphism of Hilbert spaces. As we have
mentioned, quantization on $D \times \IR$ yields the Hilbert space of a
chiral boson, depending on which boundary condition we impose. Since we
could impose either boundary condition, both chiralities are ``present.''
Perhaps a good analogy is the light-cone gauge quantization of a massless
scalar. Making one gauge choice one only sees one of two chiralities.
In the Euclidean formulation, the path integral on the bulk manifold is
equivalent to the path integral of some CFT on the boundary. Here we impose
Dirichlet boundary conditions on the gauge field and compute a wavefunctional
$\Psi(A)$ of the boundary value of $A$. It is here that we see the necessity
of both chiralities in identifying $\Psi(A)$ with a conformal field theory
partition function.

\newsec{The $AB$ theory}

\subsec{Action }

Now let us consider the $AB$ theory with action:
\eqn\abthry{
S_a =
\int {-1\over 2e_A^2} dA*dA + {-1\over 2e_B^2} dB*dB + 2\pi k A dB
}
The gauge group is $U(1) \times U(1)$, and in particular large gauge transformations
are $A \to A+ \omega^A $ and $B \to B + \omega^B$ where the $\omega^A, \omega^B$ are
closed 1-forms with integral periods.

The above treatment is asymmetric in $A,B$.
By using
\eqn\ibp{
\int AdB = \int Bd A  + \int d(BA)
}
we can convert to a theory with action:
\eqn\abthrys{
S_s =
\int {-1\over 2e_A^2} dA*dA + {-1\over 2e_B^2} dB*dB + \pi k (A dB  + B dA)
}
which is manifestly symmetric under exchanging $A\leftrightarrow B$,  $e_A \leftrightarrow e_B$.
More generally, we can use \ibp\ to formulate  the action
\eqn\abthryx{
S_x =
\int {-1\over 2e_A^2} dA*dA + {-1\over 2e_B^2} dB*dB + \pi k \bigl[ (1+x) A dB  +(1-x) B dA\bigr].
}

It is very useful to introduce $\mu:= \vert e_B/e_A\vert$
and the linear combinations
\eqn\linecombs{
\eqalign{
A^{(+)} & :={1\over \sqrt{2}}\biggl( \mu^{-1/2} B +  \mu^{1/2}  A\biggr) \cr
A^{(-)} & := {1\over \sqrt{2}}\biggl( \mu^{-1/2}B- \mu^{1/2} A \biggr) \cr}
}
the inverse relation being \invcoms.
If (and only if) $x=0$ in \abthryx\  we may use these fields to write the action as a sum of
two ``decoupled'' theories:
\eqn\apamthry{
S_s = \int \biggl[ {-1\over 2 \vert e_A e_B\vert}  d\ap* d \ap + \pi k   \ap d \ap \biggr]
+
\int \biggl[ {-1\over 2 \vert e_A e_B\vert}  d\am* d \am - \pi k   \am d \am \biggr]
}
One might conclude that the $AB$ theory is merely two copies of the one-field case with
opposite signs of $k$. {\it However, if $\mu$ is not rational then $\ap,\am$ cannot
be defined as connections on topologically nontrivial line bundles. They are not truly
independent. In particular, to implement the Gauss law on wavefunctions we cannot simply
take a product of wavefunctions for $\ap, \am$ and implement
the Gauss laws separately. This is what makes the $AB$ theory
an interesting and nontrivial extension of the one-field case.}

Another interesting new point is that the topological limit is $e_A^2 \to \infty,
e_B^2 \to \infty $ holding $\mu$ fixed. Thus, the topological sector of the theory has a
{\it continuous } parameter $\mu$, in addition to the discrete parameter $k$.
It is usually said that in the long distance limit the kinetic
terms have no effect. As we shall see, this is not quite true.
The ratio $\mu$ does affect the wavefunctions in the topological Hilbert space.


\subsec{Equations of motion}

The equations of motion are %
\eqn\diagn{
\eqalign{
d*d A^{(+)} & =  2\pi k \vert e_A e_B \vert d A^{(+)} \cr
d*d A^{(-)} & = - 2\pi k \vert e_A e_B \vert d A^{(+)} \cr}
}
and therefore there  are two  propagating vector fields of
$m^2 = (2\pi k e_A e_B)^2$.

The boundary conditions should be
such that when pulled back we have
\eqn\bcs{
-{1\over e_A^2} \delta A * dA - {1\over e_B^2} \delta B * dB + \pi k ((1+x)\delta B A + (1-x)\delta A B) =0
}

\subsec{Hamiltonian formalism: Symmetric formulation $(x=0)$ }

The Hamiltonian formulation is easily deduced by combining  \apamthry\ with section 2.3.
We have
\eqn\symmham{
S_s = -H + \int \Pi^i_+ \dot \ap_i + \Pi^i_-\dot \am_i + A_0^+ (\p_i \Pi^i_+ + \pi k \epsilon^{ij} \p_i \ap_j)  +
 A_0^- (\p_i \Pi^i_- -  \pi k \epsilon^{ij} \p_i \am_j)
}
where $H$ is the Hamiltonian (two copies of the usual one) and
\eqn\piplu{
\eqalign{
\Pi_+^i & = {1\over e_A e_B } \sqrt{g}g^{ij} (\p_0 \ap_i - \p_i \ap_0) + \pi k \epsilon^{ij} A^+_j \cr
\Pi_-^i & = {1\over e_A e_B } \sqrt{g}g^{ij} (\p_0 \am_i - \p_i \am_0) - \pi k \epsilon^{ij} A^+_j \cr}
}
The symplectic structure is
\eqn\symstr{
\Omega = \int_X  \delta \Pi_+^i \delta \ap_i
+ \delta \Pi_-^i \delta \am_i = \int_X \delta \Pi_A^i \delta A_i + \delta \Pi_B^i \delta B_i
}
The Gauss laws become:
\eqn\gausslaw{
\eqalign{
\p_i \Pi^i_B + \pi k \epsilon^{ij} \p_i A_j & = 0 \cr
\p_i \Pi^i_A + \pi k \epsilon^{ij} \p_i B_j & = 0 \cr}
}
Imposing the second class constraints of restriction to flat gauge fields gives
symplectic form
\eqn\somf{
\Omega_f = \int_X  2\pi k \epsilon^{ij} \delta B_i \wedge \delta A_j
}

Quantum mechanically, working in ``upstairs formalism'' the Gauss laws become
\eqn\newgssii{
\eqalign{
 \Psi_s(A + \omega_A, B) & = e^{-\pi i k \int \omega_A \wedge B } \Psi_s(A,B) \cr
\Psi_s(A,B+ \omega_B) & = e^{-\pi i k \int   \omega_B\wedge A  }\Psi_s(A,B) \cr}
}

Thus, if we shift by both $\omega^A,\omega^B$ then:
\eqn\newgssiii{
 \Psi_s(A + \omega^A, B+\omega^B)   = e^{-\pi i k \int \omega^A \omega^B + \omega^A B + \omega^B A } \Psi_s(A,B)
}
Note that this is only a consistent transformation law so long as $\omega^A,\omega^B$ have integral
periods and $k$ is an integer.

{\bf Remark}:   Here we encounter a truly treacherous point.
Since the action separates as in \apamthry\ one might have expected the Gauss law to be simply the
product of that for the $\ap$ and the $\am$ theory. That is, one might have expected that
\eqn\errgauss{
 \Psi_s(\ap + \op, \am + \om)  {\buildrel ? \over  = } e^{i \pi k \int( \om \am - \op \ap)} \Psi_s(\ap,\am)
}
While this indeed agrees with \newgssii\ if $\omega^A=0$ or if $\omega^B=0$ {\it it does not agree with
\newgssiii\ ! } We will discuss this subtlety more thoroughly in the general case in section 4.2 below.

\subsec{Hamiltonian analysis}

For completeness, in this subsection we give the formulation for 
an arbitrary value of
$x$. The conjugate momenta are:
\eqn\momenta{
\eqalign{
\Pi_A^i & = \tilde \Pi^i_A + \pi k (1-x) \epsilon^{ij} B_j \cr
\Pi_B^i & = \tilde \Pi^i_B + \pi k (1+x) \epsilon^{ij} A_j \cr}
}
where $\tilde \Pi^i_{A,B}$ is $x$-independent. Then
\eqn\ahmf{
\eqalign{
\int \Pi_A^i \dot A_i + \Pi_B^i \dot B_i - S & = \int H dt \cr
& +\int \Pi_A^i \p_i A_0 - \pi k (1+x) A_0 \epsilon^{ij} \p_i B_j \cr
& + \int \Pi_B^i \p_i B_0 - \pi k (1-x) B_0 \epsilon^{ij} \p_i B_j \cr}
}
Note that {\it no integration by parts has been used at this point.}

The classical Gauss law expressed in terms of $\tilde \Pi$ is $x$-independent.
On the other hand, the quantum Gauss law is
\eqn\gausslawx{
\eqalign{
 \Psi_x(A + \omega^A, B) & = e^{-i \pi (1+x)k \int \omega^A \wedge B } \Psi_x(A,B) \cr
\Psi_x(A,B+ \omega_B) & = e^{-i \pi (1-x) k \int   \omega^B\wedge A  }\Psi_x(A,B) \cr}
}

The wavefunction and Hamiltonian depend on the choice of $x$.
The general transformation between wavefunctions is
\eqn\trnsfmwv{
\Psi_x(A,B) = \Omega_x \Psi_s(A,B)= e^{-i \pi k x \int A\wedge B} \Psi_s(A,B)
}
where $x=0$ is the symmetric formulation. The Hamiltonian is obtained from
$H_x = \Omega_x H_s \Omega_x^{-1}$.

We henceforth set $x=0$ but the formulae for general $x$ can be obtained
using \trnsfmwv.

\subsec{Groundstates on $T^2$ }

The standard Hamiltonian analysis on $D\times \IR$ yields a left-
and right-moving chiral boson, once one chooses appropriate boundary
conditions. However, as in the previous section, we focus on the
Euclidean path integral on the solid torus, since the natural Dirichlet
boundary conditions on the fields distinguishes a Hamiltonian for the
singleton modes.  Therefore, we use the same
trick of considering the gauge invariant groundstate wavefunctions on $T^2$.

Again we have the factorization $\CH_f \otimes \CH_{nf}$ of the
Hilbert space and we concentrate on the wavefunctions of
flat gauge fields. We do this, and fix the small gauge
transformations by taking our wavefunctions to be  functions of the constant
gauge potentials.

The Hamiltonian can be written as:
\eqn\symmhamdens{
\eqalign{
&  H_s = -{e_A^2 \over 2 \imt }\left( {\p \over \p A_z} - 2\pi k \imt B_{\bar z} \right)
\left( {\p \over \p A_\zb} + 2\pi k \imt B_{ z} \right) \cr
&
- {e_B^2 \over 2 \imt} \left( {\p \over \p B_z} - 2\pi k \imt A_{\bar z} \right)
\left( {\p \over \p B_\zb} + 2\pi k \imt A_{ z} \right) \cr}
}
and one can solve for the Landau levels. A trick for finding these is
to write the Hamiltonian as a sum of two copies of \hamdensi, with opposite signs of $k$.
{}From \psoa\ and \psob\ we can write without further ado the wavefunctions
in the lowest Landau level (assuming $k>0$):
\eqn\prod{
\Psi_{\lambda,\bar \lambda}:=
e^{-2\pi k \imt \ap_z \ap_\zb - 2\pi k \imt \am_z \am_\zb} e^{\bar\lambda \ap_z +  \lambda \am_\zb}
}
These have  energy $2\pi {k} \vert e_A e_B \vert $ for all values
of $\lambda,\bar \lambda$, (they are not related by complex conjugation), 
 and \prod\ forms an overcomplete set
for the LLL. Again, this space of states is infinite dimensional.

Let us now follow the procedure used in the one-field case.
Averaging the wavefunctions \prod\ over the large gauge transformations
for $A,B$ to enforce the Gauss laws \newgssii\   gives a family of
gauge invariant   ground states parametrized by $\lambda,\bar \lambda$:
\eqn\averg{
\eqalign{
&
\overline{\Psi_{\lambda,\bar\lambda}} = \sum_{\omega^A, \omega^B}
\Psi_{\lambda,\bar\lambda}(A + \omega^A, B + \omega^B) \times
\cr
& \times e^{ i \pi k  \int \omega^A \omega^B
+ i \pi k  \int \omega^A B + i \pi k  \int \omega^B A }\cr}
}
Applying this to the wavefunctions \prod\ we have the averaged
sum: \eqn\solttr{ \eqalign{ &
\overline{\Psi_{\lambda,\bar\lambda}}
%
= \Psi_{\lambda,\bar\lambda}(A,B)
 \sum e^{- 2\pi k \imt  (\op_z \op_\zb + \om_z \om_\zb ) } e^{i \pi k \int \omega^A \wedge \omega^B } \cr
& e^{\bar\lambda \op_z - 4\pi k \imt  \ap_z \op_\zb - 4\pi k \imt  \am_\zb \om_z
+ \lambda \om_{\zb}  } \cr}
}
where $\omega^\pm$ are related to $\omega^A,\omega^B$ by the same linear transformation as \linecombs.

Our next move is to give an interpretation of the sum \solttr\ as an instanton sum in the partition function
of a Gaussian model on the torus. To begin, we write
\eqn\radis{
\op_z \op_\zb + \om_z \om_\zb = \mu \omega^A_z \omega^A_\zb + \mu^{-1} \omega^B_z\omega^B_\zb
}
Therefore, we see from the quadratic terms in $\omega$ in
 \solttr\  that we have {\it two Gaussian models}, one at
radius $R_A^2 = \half k \mu \alpha' $ and one at radius $R_B^2 = \half k \mu^{-1}\alpha'$.
Let us call these Gaussian fields $\phi^A$, $\phi^B$. They have periodicity
$1$ and both left- and right-movers, so $\omega^A = d \phi^A$ in an instanton
configuration on the torus. The quadratic piece of the action is
\eqn\quadrt{
S_1 = {\pi k \over 2} \int \mu d\phi^A * d \phi^A + \mu^{-1} d\phi^B * d \phi^B
}
There is evidentally a $B$-field:
\eqn\beefld{
S_2  = i \pi k \int d\phi^A \wedge d \phi^B
}

Now let us consider the coupling to the external gauge field.
Let us form the linear combinations:
\eqn\linecombsi{
\eqalign{
\phi^{(+)} & :={1\over \sqrt{2}}( \mu^{-1/2} \phi^B +  \mu^{1/2}  \phi^A)\cr
\phi^{(-)} & := {1\over \sqrt{2}}( \mu^{-1/2} \phi^B -  \mu^{1/2}  \phi^A) \cr}
}
and similarly for $\omega^{(\pm)}$.
Comparing with \solttr\   we see that the only couplings of the Gaussian fields are
$A^{(-)}_\zb$ couples to $\p_z \phi^{(-)} $ while
$A^{(+)}_z$ couples to $\p_\zb \phi^{(+)} $. To be more precise, the linear
terms correspond to an action:
\eqn\linact{
S_3 = 2 \pi i k \int \bigl[ (A^{(-)})^{0,1} \wedge \p \phi^{(-)} - (A^{+})^{1,0} \wedge \pb \phi^{(+)}\bigr]
}

{\it Thus, the rightmoving part of $\phi^+$ and the leftmoving part of $\phi^-$ couple to
the external gauge fields, and correspondingly,  one chirality  of each of $\phi^A$ and $\phi^B$ ``decouples''
from the gauge fields, but not from the metric.}

Note, that unless $\mu$ is {\it rational} the scalar fields $\phi^{(\pm)}$
do not individually have a discrete periodicity, that is, we cannot consider $\phi^+$ to
be a well-defined periodic scalar field on its own. The unusual and interesting point
is that, nevertheless $\phi^-_L + \phi^+_R$ and $\phi^+_L + \phi^-_R$ {\it are} very nearly
well-defined periodic scalars.

{\bf Remarks:}

\item{1.} Notice that (with $\alpha'=2$) the radii satisfy \rarb.
The second equation relating  $R_A$ and $R_B$ is analogous to the
$T$-duality relation. Standard $T$-duality is $R_A R_B = 2$ in units $\alpha'=2$.

\item{2.} Note that the wavefunction \prod\  only depends on $e_A, e_B$ through
the ratio $\mu$. Thus, if $e_A, e_B \to \infty$ holding $\mu$ fixed
then the wavefunction has a smooth limit. This is the limit in which
we expect the topological theory to dominate.
The gap to the next Landau level is $\sim 2\pi k \vert e_A e_B\vert$.


\subsec{Vector space of wavefunctions on $T^2$ }

At this point we could proceed with standard quantization of the
CFT defined by \quadrt\ and \linact. Let us stress that for generic
$\mu$ this conformal field theory is {\it not} a rational conformal field theory.
Nevertheless, as we will show momentarily, the space of wavefunctions 
\solttr\  spanned by $\lambda, \bar \lambda \in \IC$ is {\it finite dimensional} 
and defines an analog of the space of conformal blocks. Moreover,  
 we will show that the partition function can be written as
a finite sum of factorized terms in a fashion very reminiscent of RCFT.

In order to get at the spectrum
we will take a shortcut and simply
perform a Poisson resummation of the instanton sum \solttr. We rewrite the sum
in terms of $\omega^A,\omega^B$. We write
$\omega^A = n_1 d \sigma^1 + n_2 d \sigma^2$ and
$\omega^B = \tl n_1 d \sigma^1 + \tl n_2 d \sigma^2$. Next we do a Poisson
resummation on $n_2, \tl n_2$ and convert the sum to a form where we recognize the Hamiltonian formalism (of the conformal
field theory).   After some algebra one arrives at the result:
\eqn\narsum{
\eqalign{
\overline{\bar \Psi_{\lambda,\bar\lambda}} & ={2\imt \over k}e^{-   Q}
\sum q^{\half (p_L^2 + \tl p_L^2) } \bar q^{\half (p_R^2 + \tl p_R^2)}\cr
& \exp[ - 4\pi \sqrt{k}\imt A_z^{(+)} (p_R + \tl p_R)/\sqrt{2}
 -   4\pi \sqrt{k}\imt  A_\zb^{(-)} (p_L -  \tl p_L)/\sqrt{2} \cr
&  -
{\lambda \over \sqrt{k} }  (p_R - \tilde p_R)/\sqrt{2} - {\bar\lambda \over \sqrt{k} }   (p_L + \tilde p_L)/\sqrt{2} ] \cr}
}
where the prefactor $e^{-Q}$  is determined by
\eqn\quillen{
\eqalign{
Q & = 2\pi k \imt [ A_z^+ A_\zb^+  + A_z^- A_\zb^- ]\cr
& + 2\pi k \imt \biggl( (A_z^+)^2+ (A_\zb^-)^2 \biggr) + {1\over 8\pi k \imt } (\lambda^2
+ \bar \lambda^2) \cr} }
Now we have
\eqn\plprasym{
\eqalign{
{p_L - \tl p_L \over \sqrt{2}} & = {1\over R} m_2 - {R\over 2k } \tl m_2 \cr
{p_R+ \tl p_R \over \sqrt{2}} & = {1\over R} m_2 + {R\over 2k } \tl m_2 \cr
{p_L + \tl p_L \over \sqrt{2}} & = {1\over R} (m_2-k\tl n_1) + {R\over 2k }( \tl m_2-kn_1)  \cr
{p_R  - \tl p_R \over \sqrt{2}} & = {1\over R} (m_2- k \tl n_1)-  {R\over 2k }( \tl m_2-k n_1) \cr}
}
where $R = \sqrt{2}R_A = \sqrt{2\mu k }$, $m_2, \tl m_2, n_1, \tl n_1\in \IZ$.

At this point we can recognize the following. The sum \narsum\ is a sum over a signature $(2,2)$ Narain
lattice. We can define two sublattices: $\Lambda^A$ is the lattice of vectors ``coupling only to
$A$ and not to $\lambda$.'' Thus, it is defined by $p_R - \tl p_R =0, p_L + \tl p_L=0$. Similarly,
$\Lambda^\lambda$ is the lattice of vectors $p_R+\tl p_R =0, p_L - \tl p_L=0$. The main observation is
that these are each sublattices of signature $(1,1)$ and  $\Lambda^A \oplus \Lambda^\lambda$ is
of {\it finite index} in the full Narain lattice. The analog of the chiral splitting of RCFT is
obtained by summing over the lattice vectors in  $\Lambda^A \oplus \Lambda^\lambda$. This sum
is a factorized product of a function of $A$ and a function of $\lambda$. Then, the full sum is
given by a sum of this factorized form over the coset representatives and takes the form
\eqn\asymwvfn{
\sum_{\beta \in \Lambda^*/\Lambda} \Psi_\beta(A) \Psi_{\bar \beta}(\lambda)
}
where the lattice $\Lambda$ will be defined presently. In this way we
have defined a factorization into ``nonholomorphic conformal blocks.''

Let us make all this explicit. Note that we may write
\eqn\decouple{
\eqalign{
m_2 & = k a + \rho \cr
\tl m_2 & = k b + \tl \rho\cr
m_2 - k \tl n_1 & = k c + \rho \cr
\tl m_2 - k n_1 & = k d + \tl \rho\cr}
}
with $a,b,c,d\in \IZ$ and $0 \leq \rho,\tl \rho \leq k-1$ all uncorrelated.
In this parametrization we may write
\eqn\latoe{
\eqalign{
({p_L - \tl p_L \over \sqrt{2}};{p_R+ \tl p_R \over \sqrt{2}} ) & = a \sqrt{k\over 2\mu} e_0 - b \sqrt{\mu k \over 2} f_0 + \beta \cr
& \equiv a e_1 - b f_1 + \beta \cr
({p_L + \tl p_L \over \sqrt{2}};{p_R - \tl p_R \over \sqrt{2}} ) & = c\sqrt{k\over 2\mu} e_0 +d  \sqrt{\mu k \over 2} f_0 + \bar\beta \cr
& = c e_1 - d f_1 + \bar \beta}
}
where  $\beta = \rho/k e_1 - \tl \rho/k f_1$ and
$\bar \beta = \rho/k e_1 + \tl \rho/k f_1$. Here $e_0 := (1;1), f_0:=(1;-1)$ generate the lattice $\sqrt{2} II^{1,1}$.
The vectors $e_1,f_1$ generate a lattice
$\Lambda = e_1 \IZ + f_1 \IZ \cong \sqrt{k} II^{1,1}$. Note that $\Lambda^* \cong {1\over k } \Lambda$,
and we may regard $\beta, \bar \beta$ as representatives of elements of the dual quotient
group $\Lambda^*/\Lambda$.

Now, with any lattice of indefinite signature, but with a projection into definite signature
subspaces one may form a Siegel-Narain theta function.
The definition is reviewed in appendix A.
We may write our analogs of ``conformal blocks''
in terms of   Siegel-Narain theta functions
for $\Lambda$. Specifically, we have
%
%
%
\eqn\asymii{
\Psi_\beta(A) = \CN {2\imt \over k}
e^{-\pi k \int [\ap*\ap + \am * \am] }
\Theta_{\Lambda}(\tau,0,\beta;P;\xi(A))
}
$\CN$ is a normalization constant and
\eqn\asymvfw{
\Psi_{\bar\beta}(\lambda) =
 \Theta_{\Lambda}(\tau,0,\bar\beta;P;\xi(\lambda))
}
where we have defined:
\eqn\xisone{
\xi(A) = (\sqrt{k} 2i \imt \am_\zb ; - \sqrt{k} 2i \imt \ap_z)
}
\eqn\xiseotn{
\xi(\lambda) = ( - {\bar \lambda \over 2\pi i \sqrt{k}}; { \lambda \over 2\pi i \sqrt{k}})
}

One can now compute that
\eqn\orthowav{
\int_0^1 dA_1 dA_2 d B_1 d B_2 (\Psi_\beta(A,B))^* \Psi_{\beta'}(A,B) =
\delta_{\beta,\beta'}
{2\imt \over k^3} \vert \CN \vert^2
}
Taking into account the Jacobian factor $\imt \vert \eta \vert^4$
for going from the wavefunctional $\Psi(A(z))$ to
the wavefunction on harmonic 1-forms we finally get
\foot{Of course, we have made a choice of factorization of 
 $\imt \vert \eta \vert^4$.
Our choice was to take the positive square root.
This seems reasonable, and gives a nice representation
of the modular group below, but should be better justified.
It is certainly necessary to match to the topological theory.}

\eqn\asymii{
\Psi_\beta(A) = {2 k^{1/2}  \over \eta \bar \eta }
e^{-\pi k \int [\ap*\ap + \am * \am] }
\Theta_{\Lambda}(\tau,0,\beta;P;\xi(A))
}

It is now straightforward to compute the representation of
$\Psi_\beta(\tau,A)$ under the action of the modular group.
Specifically, the matrix elements of the $T$- and $S$-transformations
are given by
\eqn\tee{
T_{\beta, \beta'} = e^{i \pi (\beta, \beta)} \delta_{\beta, \beta'}
}
and
\eqn\ess{
S_{\beta, \beta'} = {1\over k} e^{-2\pi i (\beta,\beta')}
}
where $\beta,\beta' \in \Lambda^*/\Lambda \cong (\IZ/k\IZ)^2$ inherits a
quadratic form from the hyperbolic inner product.

This is the same representation of $SL(2,\IZ)$ as that studied in \WittenWY, and
for similar reasons. There is a natural action of the modular group on the
irrep of the discrete Heisenberg group which is a central extension of
 $H^1(X; \IZ/k\IZ ) \times H^1(X; \IZ/k\IZ)$.


\subsec{Comment on a clash of terminology}

The term ``level $k$ $U(1)$ current algebra''  is, regrettably,
used in two very different
ways in the context of the theories discussed in this paper. 
In \refs{\kutasovseiberg,\LarsenUK} Kutasov and Seiberg, and Larsen and Martinec, use  
it to refer to the structure of conformal weights $h\sim p^2/k, \tl h\sim \tl p^2/k$
where $(p,\tl p)$ lie in a (Narain) lattice of charges.  Unfortunately, the
same terminology is used with a different meaning in a closely related context in
 rational conformal field theory.
In the latter setting ``level $k$ $U(1)$ current algebra''
is the chiral algebra of the RCFT one obtains for a Gaussian
model on a rational square-radius, as described near \uoneext\ above.
One of our motivations in this paper is to clarify the relation between the
two uses of this term. We do this in the present section.

Let us consider only the momentum coupling to $\ap,\am$.
Let us define the left and right ``charges''  by:
\eqn\charges{
\eqalign{
u_L & :=\xi (p_L - \tl p_L) \cr
u_R & :=\xi  (p_R + \tl p_R) \cr}
}
where $\xi$ is a real normalization constant to be determined below.

The set of charges \charges\ forms a lattice in $\IR^{1,1}$ defined by
\eqn\chargelattice{
\Lambda:= \{ (u_L; u_R)\vert n,m,\tl n,\tl m\in \IZ  \} \subset \IR^{1,1}
}
  This lattice is generated by
integral combinations of 2 vectors:
\eqn\fourvect{
\eqalign{
e_1 & = \xi {1\over R_A}  e_0  \cr
f_1 & = \xi {R_A\over k}  f_0  \cr
}
}
where $e_0 := (1;1), f_0:=(1;-1)$ generate the lattice $\sqrt{2} II^{1,1}$.
Thus, $e_1\cdot f_1 = 2\xi^2/k$, while $e_1^2= f_1^2 =0$.
The charge lattice is $e_1 \IZ \oplus f_1 \IZ$.
So choosing
\eqn\keven{
\xi = \sqrt{k \over 2}
}
we obtain a self-dual lattice.

In terms of these charges we can write the conformal weights of the
states counted in \narsum\ as:
\eqn\leftconf{
\eqalign{
h& = \half (p_L^2 + \tl p_L^2)\cr
& = {1\over 4} \bigl( p_L + \tl p_L  \bigr)^2  + {1\over 2k }  u_L^2  \cr}
}
\eqn\rightconf{
\eqalign{
\tl h& = \half (p_R^2 + \tl p_R^2)\cr
& = {1\over 4} \bigl( p_R - \tl p_R  \bigr)^2  + {1\over 2k }  u_R^2  \cr}
}

Now, for fixed values of the ``spectator charges'' $(p_L + \tl p_L; p_R - \tl p_R)$
we recognize, after using \keven\
that the dependence of the conformal weight on $(u_L; u_R)$
is that of ``level $k$ $U(1)$ current algebra.''  Note especially that
\eqn\comparandy{
\eqalign{
{1\over 4} (p_L- \tilde p_L)^2 - {1\over 4}(p_R + \tl p_R)^2 & = {1\over 4\xi^2} (u_L^2- u_R^2) \cr
& = - {m_2\tilde m_2\over k} = {N\over k}  \cr}}
where $N$ can be any integer.

\bigskip 
{\bf Remark.}{\it The purely topological quantization.} In  \WittenYA\
Witten studied the off-diagonal Chern-Simons theory for  the case that   $k=1$  and concluded
that the pure Chern-Simons theroy is ``trivial.''
%
%
%
It is straightforward to analyze the
 purely topological theory on $D \times \IR$  using the methods of \WittenHF\MooreYH\ElitzurNR.
One finds a left and a right-moving boson, but, we stress, {\it these are not
the left- and right-moving components of a single boson of well-defined
discrete periodicity.}
One can compute  $L_0 - \bar L_0$ in this approach and one finds $L_0 - \bar L_0 = N/k$.
Without further input it is difficult to decide whether we should allow all integers $N$,
or whether one should project to $N=0~\mod~ k$.
The approach we are taking in this paper 
answers that question. We see that  the integer $N$ in \comparandy\
can   be {\it  any} integer.

\newsec{General massive abelian Chern-Simons  theories}

Both in the theory of the quantum hall effect \refs{\read, \BlokAN,\FrohlichAK,\FrohlichQM,\FrohlichVQ, \zeereview,\moorewen}
and in $AdS_3 \times \S^3 \times T^4$
one is naturally led to wonder about the extension of the above remarks to a collection of
abelian gauge fields $A^\alpha$, $\a = 1,\dots, d$.
We take the  action
\eqn\manyflds{
\int - {1\over 2 e^2}  \lambda^{-1}_{\a\b} dA^\a * dA^\b + 2\pi K_{\a\b}   A^\alpha dA^\beta
}
and the gauge fields are normalized so that  $F^\alpha$ has integral periods.
The gauge group is $U(1)^d$.
The Euclidean version is $e^{-S_E}$ with
\eqn\manyflds{
\int  {1\over 2 e^2}  \lambda^{-1}_{\a\b} dA^\a * dA^\b - 2\pi i K_{\a\b}   A^\alpha dA^\beta
}
The coupling $e^2$ has dimensions of mass, while
  $\lambda^{-1}_{\a\b}$ is a dimensionless positive definite symmetric matrix.
Without loss of generality
we may assume it has fixed determinant, say determinant one.

\lref\DijkgraafPZ{
R.~Dijkgraaf and E.~Witten,
``Topological Gauge Theories And Group Cohomology,''
Commun.\ Math.\ Phys.\  {\bf 129}, 393 (1990).
}

We will assume that $K_{\a\b}$ is nondegenerate. As we have seen above,
 by adding total derivatives, we can assume that $K_{\a\b}$ is symmetric,
and these total derivatives do not affect the quantization of the theory.
In order that the
action makes sense on arbitrary manifolds we must have
\eqn\constr{
\int_{M_4} K_{\a\b} c_1^\a c_1^\b \in \IZ
}
where $c_1^\a$ is a vector of integer cohomology classes on the four-manifold 
$M_4$. Clearly $K_{\a\a}\in \IZ$.
Using the example of  $\S^2 \times \S^2 $ we see that $K_{\a\b}+ K_{\b \a} \in \IZ$
for $\a\not= \b$, and this is sufficient for well-definedness
in general. \foot{By choosing
a spin structure and only considering bounding manifolds
compatible with the spin structure we can allow theories with more
general Chern-Simons couplings \DijkgraafPZ. This involves several new issues, and we will
not investigate that case here.}
Thus, we
conclude that $2K_{\a\b}$ is a nondegenerate,   even, integral, symmetric matrix.
It can have any signature. This matrix defines an integral lattice which we denote
$\bar \L$.  We will denote the integral lattice generated by $-2\kab$ by $\L$.

The matrix of Chern-Simons couplings $\lambda^{-1}_{\a\b}$ has inverse
$\lambda^{\a\b}$. The topological limit is obtained by taking $e^2\to \infty$.
Thus we expect both $\lambda^{\a\b}$ and $\kab$ to show up in constructing
the wavefunctions for the topological Hilbert space.

%

\subsec{Quantization of the purely topological theory}

The quantization of the pure
Chern-Simons theory is completely straightforward and was
in fact already analyzed to some extent in \ElitzurNR. We have
\eqn\topps{
[A^\a_j, A^\b_k] = {\epsilon_{jk} K^{\a\b} \over 2\pi i }
}
Choosing a real polarization on the torus we have wavefunctions $\Psi(A_1^\a)$.
Implementing the Gauss law for transformations in the $\sigma^2$ direction
we find the wavefunctions   are supported on gauge potentials such
that $\kab A_1^\b \omega_2^\a \in \IZ$,
that is, on points in $\L^*$. The Gauss law for transformations in the
$\sigma^1$ direction shows that the wavefunction descends to $\L^*/\L$.
This leads to a standard representation of a finite Heisenberg group, and
is associated to a representation of $SL(2,\IZ)$ in a natural way.


\subsec{Hamiltonian analysis and Gauss law}

The conjugate momentum is
\eqn\conjm{
\Pi^i_\a = \tl \Pi^i_\a + 2\pi K_{\b\a}\epsilon^{ij}A^\b_j
}
where $\tl \Pi^i_a = \lambda^{-1}_{\a\b} g^{ij} \sqrt{g} (\p_0 A^\b_j - \p_j A^\b_0) $
is the electric field. The Hamiltonian is
\eqn\hamdsn{
H = \int_X {g_{ij} \over 2\sqrt{g}} \lambda^{\a\b} \tl \Pi^i_\a \tl \Pi^j_\b
+ \half \lambda^{-1}_{\a\b} F^\a*_2 F^\b
}

The classical Gauss law is
\eqn\glcss{
\p_i \Pi^i_\a + 2\pi \kab \p_i A^\b_j \epsilon^{ij} =0
}

Implementing the  quantum Gauss law one encounters a subtlety.
Let $\omega^\a$ be a 1-form with integral periods. Define the operator
\eqn\qgauss{
\CG_\a(\omega^\a) :=  i \int \omega_i^\a \Pi^i_\a + 2\pi \sum_\b \omega^\a\kab A^\b
}
where {\it there is no sum on $\a$}. One easily computes that
\eqn\expd{
e^{\CG_\a(\omega^a)} e^{\CG_\a(\tl \omega^\a)} = e^{\CG_\a(\omega^\a+\tl \omega^\a)+ 2\pi i \int K_{\a\a}
\omega^\a \tl \omega^\a } = e^{\CG_\a(\omega^a+\tl \omega^a)}
}
since $K_{\a\a}$ is integral. Similarly, if $\a\not=\b$ then
\eqn\expd{
e^{\CG_\a(\omega^\a)} e^{\CG_\b( \omega^\b)} = e^{\CG_\a(\omega^\a) + \CG_\b( \omega^\b)- 2\pi i \int K_{\a\b}
\omega^\a  \omega^\b }  = e^{\CG_\b( \omega^\b)}  e^{\CG_\a(\omega^\a)}
}
Since $\kab\in \half \IZ$, the operators $e^{\CG_\a}$ are simultaneously commuting and
can all be imposed as constraints. However, one {\it cannot} enforce the Gauss laws
\eqn\cantdo{
e^{\CG_\a(\omega^\a) + \CG_\b( \omega^\b)}\Psi = \Psi
}
because they have a nontrivial cocycle in the group law. This is the 
origin of the $B$-field $(4.22)$ in the holographically dual theory. 

Enforcing all the Gauss laws $e^{\CG_\a} \Psi = \Psi $ for $\a=1,\dots, d$ is equivalent to the
quantum Gauss law:
\eqn\quags{
\Psi(A^1+ \omega^1, \dots, A^d + \omega^d) =
e^{-2\pi i \int \sum_{\a<\b} \kab \omega^\a \omega^\b}
e^{-2\pi i \int \sum_{\a, \b} \kab \omega^\a A^\b}
\Psi (A^1 , \dots, A^d  )
}
%


\subsec{Landau levels}

On the flat torus we have Hamiltonian
\eqn\hamdtor{
H = - \int {1\over 2\imt} \lambda^{\a\b } \biggl( {\p \over \p A^\a_z } - 4\pi \imt
K_{\g\a} A^\g_\zb\biggr)
\biggl( {\p \over \p A^\b_\zb } + 4\pi \imt K_{\g\b} A^\g_z \biggr)
}
where we have chosen a normal ordering. On the plane a complete set of
functions for  the lowest Landau level is generated by the wavefunctions

\eqn\lllhr{
\Psi_{v,\bar v} = \exp\biggl[ - 4\pi \imt \tmab A^\a_z A^\b_\zb + \bar v_\a A^\a_z + v_\a A^\a_\zb \biggr]
}
where $\bar v_\a, v_\a$ are independent complex vectors. 

One finds that \lllhr\ is an eigenfunction of \hamdtor\ if and only if
\eqn\evconsd{
\eqalign{
[\lambda K, \lambda \mu] & = 0 \cr
(\lambda \mu)^2 & = (\lambda K)^2 \cr
(\mu + K)\lambda v & = 0 \cr
(\mu - K) \lambda \bar v& =0 \cr}
}
where for simplicity we have assumed that $\tmab$ is symmetric.

Now, for normalizable wavefunctions we want $\tmab$ to be positive hermitian.
In this case, the last two equations in \evconsd\ involve projection operators.
Now, note that $\lambda^{1/2} K \lambda^{1/2}$ is a symmetric form and therefore
can be diagonalized by a real orthogonal matrix $\CO$:
\eqn\diagk{
K = \lambda^{-1/2} \CO \pmatrix{ \Delta^+ & 0 \cr 0 & \Delta^-\cr} \CO^{tr} \lambda^{-1/2}
}
where $\Delta^\pm$ are diagonal matrices with $\Delta^+_{ii} >0$ and $\Delta^-_{ii}<0$.
We therefore can solve our equations by letting
\eqn\diagm{
\mu = \lambda^{-1/2} \CO \pmatrix{ \Delta^+ & 0 \cr 0 & - \Delta^-\cr} \CO^{tr} \lambda^{-1/2}
}
Thus, $\mu$ is positive definite.
The energy eigenvalue with our normal ordering is $- 8 \pi \sum_i \Delta^-_{ii} $.

It is useful to introduce the vector space $V\cong \IR^d$ where $A^\a$ is valued.
We can regard $\mu, K \in V^*\otimes V^*$ while $\lambda,\mu^{-1} \in V\otimes V$.
 Note that $v=v_\a , \bar v = \bar v_\a$ are valued in $V_c^*$. The subscript
$c$ means that we have complexified. 
Note that $\Gamma^{\a}_{~ \b} = \mu^{\a\g} K_{\g\b} $ is an operator
$\Gamma: V \to V$ and satisfies $\Gamma^2=1$. Here $\mu^{\a\g}\mu_{\g\b}=\delta^{\a}_{~\b}$.
 We define projection matrices
\eqn\projmtr{
P_\pm : = \half ( 1 \pm \mu^{-1}K)
}
and accordingly we have subspaces $V_\pm:= P_\pm V$. With this choice $\lambda v \in V_-$, and
$\lambda \bar v \in V_+$.  The following identities are useful.
Since $\mu$ is symmetric, $P_\pm^{tr}$ are also
projection matrices, and $\mu P_{\pm} = P_{\pm}^{tr} \mu$.
Moreover, $(\mu^{-1}K)^{tr} = \lambda^{-1} (\mu^{-1} K) \lambda$, so
$P_\pm^{tr} = \lambda^{-1} P_\pm \lambda$,  and so we can also say
that
\eqn\vtrs{
\eqalign{
v^{tr}P_+ & =0 \cr
\bar v^{tr} P_- & = 0 . \cr}
}

\subsec{Averaged wavefunction}

Now we can proceed as before with the average
\eqn\vads{
\overline{\Psi_{v,\bar v}} = \sum_{\omega^\a} \Psi_{v,\bar v}(A+\omega)
e^{2\pi i \sum_{\a<\b} \int \omega^a \kab \omega^\b + 2\pi i \int \omega^\a \kab A^\b}
}

Expanding out we find the soliton sum of a theory of bosons $\phi \in V$,
with periodicity $\phi^\a \sim \phi^\a + 1$. The action is
\eqn\bosonact{
S = 2\pi \int d \phi^\a \mu_{\a\b} * d \phi^\b - i \pi \int B_{\a\b} \omega^\a \wedge \omega^\b
}
where $B_{\a\b} = - B_{\b\a} $ is a $B$-field defined by
\eqn\befs{
B_{\a\b} = \kab \qquad\qquad \a < \b
}
The chiral coupling to the gauge fields is
\eqn\chrsple{
- 4\pi i \int \p \phi^\a \mu_{\a\b} (P_- A^{0,1})^\b
+ 4\pi i \int \pb \phi^\a \mu_{\a\b} (P_+ A^{1,0})^\b
}

Thus, only holomorphic currents valued in $V_-$ couple to $A_\zb$, while only
antiholomorphic currents valued in $V_+$ couple to $A_z$. Similarly, the coupling
to $v,\bar v$ in \vads\ is just:
\eqn\vcops{
\exp[   \bar v^{tr} P_+ \omega_z  +  v^{tr} P_- \omega_{\zb} ]
}

We stressed above in the $AB$ theory that
 $\phi^+, \phi^-$ were not scalars with definite periodicity.
The generalization of this statement is that the gauge group
(or periodicity lattice for $\phi$)  defines a lattice
$\IZ^d \subset V$. The subspaces $V_\pm$ in general do not contain any lattice vectors.
Thus, the  chiral scalars $P_- \phi_L $ and $P_+ \phi_R$ in general do not form a
single well-defined scalar. Indeed, the lattice $\bar\L$ in
general has signature $(r_+,r_-)$ with $r_+ \not= r_-$.

\subsec{Vector space of states on $T^2$}

One can quantize the theory of chiral bosons as before. The averaged wavefunction
may be expressed in terms of a sum over an even unimodular Narain lattice
of signature $(d,d)$. We endow the real vector space $V \oplus V$ with the
quadratic form:
\eqn\narsin{
(p_L; p_R)\cdot (q_L; q_R)  := p_L^\a \tmab q_L^\b - p_R^\a \tmab q_R^\b
}
Note that there are now two totally independent projections in the game.
We have $P_\pm$ projecting onto subspaces of $V$ determined by the
Chern-Simons coupings $\lambda,K$.  In addition   we have the left- and right-moving
projections of Narain theory, related to the chirality of the bosons.
 The latter projections are denoted by $L,R$.
The embedding of $II^{d,d} \otimes \IR \subset V\oplus V$ is accomplished by
the basis vectors:
\eqn\bemds{
\eqalign{
e_\a & = {1\over \sqrt{2}}(\delta^\gamma_{~\a} - \mu^{\gamma\zeta}B_{\zeta\a};
\delta^\gamma_{~\a} + \mu^{\gamma\zeta}B_{\zeta\a} ) \qquad\qquad \a = 1,\dots, d\cr
f^\a & = {1\over \sqrt{2}}(\mu^{\gamma\a}; - \mu^{\gamma\a}) \qquad\qquad \a = 1,\dots, d \cr}
}
In the above formulae we denote the components of the $L,R$ projection by the
superscript $\gamma$.
One easily checks that
\eqn\inps{
\eqalign{
e_\a \cdot e_\b & = 0 \cr
f^\a \cdot f^\b & = 0 \cr
e_a\cdot f^\b & = \delta^{\a}_{\b} \cr}
}
and hence integral combinations of these vectors define an embedding of
the even unimodular lattice $II^{d,d}$ into $V\oplus V$.

Now, by examining \vads\ or by quantizing
\bosonact\chrsple\ one finds that only the projection of $A_\zb$ into $V_-$ couples to $p_L$
while only the projection of $A_z$ into $V_+$ couples to $p_R$.
Similarly, in the averaged wavefunction, the projection of $\bar v$ into $V_+$
couples to $p_L$ while the projection of $v$ into $V_-$ couples to $p_R$.
Thus we  define two collections of $d$ vectors:
\eqn\nunubar{
\eqalign{
\nu_\a & = \sqrt{2} \biggl( (P_-)^{\gamma}_{~ \a};  (P_+)^{\gamma}_{~ \a}\biggr)\qquad \a = 1, \dots ,d  \cr
\bar \nu_\a & = \sqrt{2} \biggl( (P_+)^{\gamma}_{~ \a};  (P_-)^{\gamma}_{~ \a}\biggr)
\qquad \a = 1, \dots ,d \cr}
}
The real span of the $\nu_\a$ is a subspace of $V_L \oplus V_R$ which we can
denote $V_{-,L} \oplus V_{+,R}$ while the real span of the $\bar \nu_\a$
is $V_{+,L} \oplus V_{-,R}$.

Moreover, one  easily computes that
\eqn\nuprods{
\eqalign{
\nu_\a \cdot \nu_\b & = - 2\kab \cr
\bar \nu_\a \cdot \bar \nu_\b & = + 2\kab \cr
\nu_\a \cdot \bar\nu_\b & =0 \cr }
}
and hence integral combinations of $\nu_\a$ generate a lattice $\L$, while
integral combinations of $\bar \nu_\a$ generate a lattice $\bar \L$. Furthermore,
\eqn\nupfrs{
\eqalign{
f^\beta\cdot \nu_\a & = \delta^{\beta}_{~\a} \cr
f^\beta\cdot \bar \nu_\a & = \delta^{\beta}_{~\a} \cr
e_\a \cdot \nu_\b & = - \kab + B_{\a\b} \cr
e_\a \cdot \bar \nu_\b & =  \kab + B_{\a\b} \cr}
}%
Since $II^{d,d}$ is unimodular,  $\L$ and $\bar \L$ are sublattices of  the Narain
lattice generated by $e_\a, f^\a$.
The lattice $\L \oplus \bar \L$ is of
finite index in $II^{d,d}$. We can now uniquely decompose any  Narain vector in terms
of its projection into $\L \otimes \IR \oplus \bar \L \otimes \IR$. These projections
consist of a vector in $\L$ plus a glue vector in $\L^*/\L$. To be specific, there exist
a finite set of vectors $\beta \in \L^*$, $\bar \beta \in \bar \L^*$ such that
$\beta + \bar \beta \in II^{d,d}$ and such that we can write:
\eqn\decompso{
p = n^\a e_\a + m_\a f^\a = p_\L + p_{\bar \L}
}
where
\eqn\dcoms{
\eqalign{
p_\L & = (\ell^\a  -  \half K^{\a\b}\delta_\b ) \nu_\a = \ell^\a  \nu_\a + \beta  \cr
p_{\bar \L} & = (\bar\ell^\a  +  \half K^{\a\b}\delta_\b ) \bar \nu_\a
= \bar\ell^\a \bar\nu_\a  +  \bar \beta \cr}
}
Here $\ell^\a,\bar \ell^\a$  are independent vectors of integers.
Moreover,   $\delta_\a$ runs over a finite set of
 integral vectors. Put differently, we can make a 1-1 transform on the integers
$n^\a,m_\a$ in \decompso\ in such a way that and we use a finite set
of vectors $\delta_\a$ representing $\L^*/\L$.
To be specific, every vector of integers $m_\a$ can be uniquely
written in terms of a vector of integers $\ell^\a$ and the vectors $\delta_\a$ as
\eqn\dfsne{
m_\a = 2\kab \ell^\b + \delta_\a
}
We may take $\beta  = -  \half K^{\a\b}\delta_\b \nu_\a $ and
$\bar \beta  = +  \half K^{\a\b}\delta_\b \bar \nu_\a $.  The mapping $\beta \to \bar \beta$
should be viewed as an isomorphism of dual quotient groups $\L^*/\L \to \bar\L^*/\bar \L$.
Indeed, the Nikulin embedding theorem \nikulin\ describes the embedding of an even integral lattice,
such as $\L$, into any even unimodular lattice, such as $II^{d,d}$,
 in terms of an isomorphism of dual quotient groups between $\L$ and its complementary lattice
$\bar \L$.  Here we have made that isomorphism explicit.

Now, it turns out that the left- and right- projections to $p_L$, $p_R$  are compatible
with the projections $P_\pm$ onto the subspaces $V_\pm$. Thus, for example, we have
\eqn\commtprj{
\nu_{\a,L} \cdot \bar \nu_{\b, L} = 0 \qquad \nu_{\a,R}\cdot \bar \nu_{\b,R} =0
}
Thus, we can split the sum over the Narain lattice $II^{d,d}$ into a finite
sum over $\L^*/\L$ of factorized wavefunctions coupling only to $A$ and $v,\bar v$,
respectively. The averaged wavefunction can be written in terms of higher-level
Siegel-Narain theta functions as:
\eqn\vinfal{
\overline{\Psi_{v,\bar v}} = e^{-2\pi \int \tmab A^\a * A^\b} {\imt^{d/2} \over \sqrt{\det \mu}}
\sum_{\beta\in \L^*/\L}
\Theta_\L(\tau,0,\beta;P; \xi(A)) \Theta_{\bar \L} (\tau,0,\bar \beta;P; \xi(v))
}
where
\eqn\xia{
\xi(A) = - \sqrt{8} \biggl( P_- (i \imt A_\zb) ; P_+ (i \imt A_z) \biggr)
}
\eqn\xiv{
\xi(v) = {\sqrt{2}\over 2\pi i }  \biggl( P_+ (\mu^{-1}\bar v) ; - P_- (\mu^{-1} v) \biggr)
}

As in the previous case, \vinfal\ only gives the wavefunctional of the gauge fields
 up to a normalization
constant. As before, a basis of wavefunctions for the topological theory
can be given in the form
\eqn\basis{
\Psi_\beta = e^{-2\pi \int \tmab A^\a * A^\b} {\Theta_\L(\tau,0,\beta;P; \xi(A))\over \eta^{r_+}
\bar \eta^{r_-} }
}
where $(r_+,r_-)$ is the signature of $\L$. The
representation of the modular group is precisely analogous to what we had before:
\eqn\teekk{
T_{\beta, \beta'} = e^{- 2\pi i (r_+-r_-)/24} e^{i \pi (\beta, \beta)}
\delta_{\beta, \beta'}
}
%
\eqn\esskk{
S_{\beta, \beta'} = {1\over \sqrt{\vert \L^*/\L\vert}} e^{-2\pi i (\beta,\beta')}
}
where $ \Lambda^*/\Lambda  $ inherits a
quadratic form defined by
\eqn\inqf{
q(\beta ~\mod~ \L) := (\beta, \beta) ~\mod~ 2,
}
where $\beta$ is any lift of $\beta ~\mod~ \L$ to a vector in $\L^*$.

{\bf Remarks}:

\item{1.} We can say precisely in what sense this is a generalization of the
chiral splitting of RCFT. The latter case corresponds to the case where
$\L$ and $\bar \L$ are lattices of definite signature, hence $\L$ is purely
left-moving and $\bar \L$ is purely right-moving. We would like to stress that,
despite the notation, $\Theta_\Lambda(\tau,\cdots)$ is {\it not} holomorphic
in $\tau$ if $\Lambda$ is not of definite signature.

\item{2.} In \moorewen\
there are some related computations. However, these authors assume that the
edge state bosons have well-defined periodicity, and hence are not describing
the dual to the most general abelian Chern-Simons theory.


\subsec{Generalization to higher genus surfaces}

The above computations generalize to higher genus surfaces $X$. Our wavefunctions 
are functions on the vector space of real harmonic one-forms on $X$. 
We define coordinates by choosing a basis $\omega^a = \omega^a_z dz$ of holomorphic 
1-forms, while $\bar \omega^{\bar a} $ is a basis of anti-holomorphic 1-forms, 
with $a,\bar a = 1,\dots h$, $h$ is the genus of $X$. 

Recall that the momentum is a vector-valued density, so 
\eqn\coords{
\Pi^i_\a {\p \over \p x^i} \otimes d^2 x , \qquad \a = 1, \dots, d
}
is coordinate invariant. The Hamiltonian   is
\eqn\hamdsn{
H = \int_X {g_{ij} \over 2\sqrt{g}} \lambda^{\a\b} \tl \Pi^i_\a \tl \Pi^j_\b  d^2 x + \half \lambda^{-1}_{\a\b} F^\a*_2 F^\b 
}

Our phase space is the cotangent space $T^* \Gamma(\Omega^1(X) )$. Our strategy is to restrict 
to the sub-phase space of the cotangent bundle to the space of real harmonic forms. We 
refer to this as the ``small phase space'' for brevity. 
Just as on the torus, we can introduce complex coordinates so that 
\eqn\cplxcr{
g_{ij} d\sigma^i\otimes  d\sigma^j = g_{z\bar z} (dz \otimes d\bar z + d\bar z \otimes d z) 
}

\def\ba{{\bar a}} 
\def\bb{{\bar b}}
\def\bc{{\bar c}}
\def\bd{{\bar d}} 
\def\bz{{\bar z}} 

Now, in restricting to the small phase space we take
\eqn\coords{
A^\a = \sum_{a=1}^h \bigl(  A_a^\a \omega^a_z dz + A_\ba^\a \omega^\ba_\bz d\bz \bigr)
}
Note that $A_\ba^\a = (A_a^\a)^*$ are complex coordinates on phase space and are $z,\bz$-independent. 
The symplectic form is 
\eqn\symple{
\Omega =\int_{X} \delta \Pi_\a \wedge \delta A^\a 
}
  Restricting to the subspace \coords\
we define the conjugate coordinates on phase space by 
\eqn\defps{
\eqalign{
\delta \Pi^z_\a & = \delta \Pi^b_\a  (\tau^{-1})_{\bc b} \omega^{\bc}_\bz\cr
\delta \Pi^\bz_\a & = \delta \Pi^\bb_\a  (\tau^{-1})_{\bb c} \omega^{c}_z\cr}
}
Here we have introduced the period matrix 
\eqn\taumatrx{
\tau^{a\bc} := \int_{X} \omega^a \wedge  \omega^\bc
}
The symplectic form on the small phase space is 
\eqn\restsf{
\Omega = \delta \Pi^a_\a \wedge \delta A_a^\a + cplx. conj.
}
and this fixes the quatization: 
\eqn\quantz{
\Pi^a_\a = - i {\p \over \p A^a_\a} \qquad \Pi^\ba_\a = - i {\p \over \p A^a_\ba} 
}
There is no misprint here. We have a minus sign on the RHS of both expressions. 

Now we find 
\eqn\exprprs{
\eqalign{
\tilde \Pi^z_\a =
& = - i \omega^{\ba}_{\bz} (\tau^{-1})_{\ba b} \biggl[ {\p \over \p A^\a_b} - 2\pi K_{\b\a}i \tau^{b\bc}A^\b_\bc \biggr] \cr
\tilde \Pi^\bz_\a =& 
- i \omega^{c}_{z} (\tau^{-1})_{\bb c} \biggl[ {\p \over \p A^\a_\bb} + 2\pi K_{\b\a}i \tau^{d\bb}A^\b_d \biggr] 
\cr}
}
Finally we substitute into \hamdsn\ with $F=0$.
We get a $(1,1)$ form, and the integral over $X $ is 
\eqn\harmham{
 H =  
-\half \lambda^{\a\b} \tau^{-1}_{\bd b} \Biggl\{
\biggl({\p \over \p A_b^\a} - 2\pi i K_{\gamma\a} \tau^{b\bar e} A_{\bar e}^\gamma\biggr), 
\biggl({\p \over \p A_{\bar d}^\b} + 2\pi i  K_{\delta\b}
 \tau^{f\bar d} A_{f}^\delta \biggr)\Biggr\}
}

Thus, we see that the above discussion easily generalizes to arbitrary genus. 
Roughly speaking $\tau$ becomes the period matrix, and we replace 
$\lambda \to \lambda \otimes \tau^{-1}$ while $K \to K\otimes \tau$.

\newsec{Open problems and further questions}

The present paper will appear somewhat trivial to many readers.
While the computations are elementary --- after all we are discussing free field
theory --- we think it is important to have a clear idea of the wavefunctions
which naturally come up in the study of holography of massive Chern-Simons
theory. To conclude, we discuss briefly some natural continuations of the
above results.

First, much   of the structure of the rational Gaussian model can be understood in terms of the
extended chiral algebra, where one extends the $u(1)$ chiral algebra generated by
$i \p \phi(z)$ by the operators $e^{\pm i \sqrt{2k} \phi(z)}$. This defines
the  ``level $k$ $U(1)$ chiral algebra'' in the sense of RCFT.   The conformal blocks of the
RCFT are the holomorphic theta functions which are characters of
 this  chiral algebra. Is there an analogous nonholomorphic algebra  in the present case?
 A related question is to understand in detail how
Wilson lines piercing the cylinder/torus
correspond to vertex operator insertions in the boundary conformal field theory.

Second, there might be some interesting connections with the idea of integrable
structures in the AdS/CFT correspondence. In the above discussion we have
always assumed that $\mu$ is irrational. However, when $\mu$ is rational
the dual  {\it is} an RCFT. By the correspondence there is an infinite
set of ``extra'' holomorphic conserved charges in the string dual
on  $AdS_3 \times K_7$. It would be
worth seeing if this enhanced symmetry gives useful information
on the holographic correspondence and how, in detail,  it leads to greater solvability
of the string theory.

A natural question one can ask is what the nonabelian generalization of the $AB$-type
theory might be. In fact, Kaluza-Klein reduction of six-dimensional supergravity
on $AdS_3 \times \S^3$ yields a very interesting and subtle generalization of
$SU(2)$ massive Chern-Simons theory, which deserves to be understood better than
it is at present \refs{\DKSS,\LPS,\LPSb,\APT,\Mathur,\NS}.

The simple free field theory we are discussing might offer a useful laboratory
to explore some issues of holography. In the massive Chern-Simons theory, which
is not holographic, there is a many-to-one map from ``interior data'' such as the
choice of metric within the solid torus, or the presence of local operators, to
the coefficients $\zeta^\beta$ of the wavefunction appearing in \zeelin. Some aspects of this map
(such as the metric dependence) could in principle be made quite explicit.
When embedded in string theory the analogous $\zeta^{\beta}_{\rm string}$ 
in \zeestring\ is supposed to
be a ``1-1 map'' between the internal data and the data of the boundary conditions
of all the string fields.   Understanding this better, in the present context
might be  useful in addressing the
puzzles raised in the recent paper \MaldacenaRF. Let $X_3 = \IH^3/\Gamma$
be the quotient of hyperbolic 3-space by a quasi-Fuchsian group. Then
there are Riemann surfaces $X, X'$ at the two ends. The
partition function of the massive abelian Chern-Simons theory on
this manifold has the ``entangled'' form:
\eqn\entangl{
\sum_{\beta,\beta'} \zeta^{\beta\beta'} \Psi_\beta(A) \Psi_{\beta'}(A')
}
where $\zeta^{\beta\beta'} $ depends on the details of what
operators have been inserted in the interior of the 3-manifold.
According to our general conjecture, $\zeta_{\rm string}^{\beta\beta'}$
should only depend on (arbitrary) boundary conditions on the two
end surfaces $X, X'$.
AdS/CFT leads us to expect that it is  an outer product
of two vectors. We see no {\it a priori} reason why this cannot be
true, and we believe this is the resolution of the puzzles described in
\MaldacenaRF.

It is quite natural to try to extend the discussion here to
two higher dimensional analog systems.
The first natural generalization is to the  $(B_{NS},B_{RR})$
system on spacetimes which are asymptotically hyperbolic and 
have boundary $X_4 \times \S^5$, where $X_4$ is a 4-manifold. 
The analysis of the associated topological field theory 
was undertaken in \WittenWY.  
In \WittenWY\ the kinetic terms were neglected, as is
appropriate for the study of the representation of $SL(2,\IZ)$.
However, we have seen that for  finer questions involving natural bases of
wavefunctions one should retain the kinetic terms.  
A computation analogous to that above
indeed produces the partition function of a boundary theory of
a $U(1)$ gauge field coupling to a ``chiral'' combination of
$(B_{NS},B_{RR})$.  In this case,  the ``new'' parameters, analogous to $\mu$ above, 
include the complex dilaton $\tau$ of the type IIB string and the conformal 
class of the metric on $X_4$.   We expect that
the full string theory partition function gives
an analog of the decomposition \zeestring, where $\zeta^\beta_{\rm string}$
is the partition function of $SU(N)/\IZ_N$ SYM theory in different
't Hooft flux sectors,  and
$Z_{\rm string}$ is the partition function of $U(N)$ SYM theory.

Finally, we hope that  the method of this paper will
 help in understanding better the
pairing   between the
5-brane partition function and the supergravity path integral
for the $C$-field and that there will be a nice combination
of the   results of \DiaconescuBM\ with the techniques of this
paper.


\bigskip
\noindent{\bf Acknowledgements:}

We would like to thank J. Maldacena and  E. Witten for discussions
and  C. Schweigert and N. Read for correspondence.
G.M. would like to that the KITP for hospitality during
the course of some of this work.
This work was conducted during the period S.G. served as a Clay
Mathematics Institute Long-Term Prize Fellow. The work of E.M. is
supported in part by DOE grant DE-FG02-90ER40560, that of G.M. by
DOE grant DE-FG02-96ER40949 and that of A.S. by DE-FG02-91ER40654.

\appendix{A}{Siegel-Narain  Theta functions}

Let $\Lambda$ be a lattice of signature $(b_+,b_-)$.  Let $P$
be a decomposition
of $\Lambda\otimes \IR$ as a sum of orthogonal subspaces
 of definite
signature:
\eqn\dfsign{ P:\Lambda \otimes \IR \cong
\IR^{b_+,0} \perp \IR^{0,b_-}
}
Let $P_\pm(\lambda)= \lambda_\pm $ denote the projections onto the two
factors.
We also write $\lambda = \lambda_+ + \lambda_-$.
 With our conventions $P_-(\lambda)^2 \leq 0 $.

Let  $\Lambda+ \gamma $ denote a translate of the lattice
$\Lambda$.
We define the Siegel-Narain theta function
\eqn\sglthet{
\eqalign{
\Theta_{\Lambda + \gamma} (\tau, \alpha,\beta; P, \xi)
\equiv
&
 \exp[{ \pi \over  2 y} ( \xi_+^2 - \xi_-^2) ] \cr
\sum_{\lambda\in \Lambda + \gamma}
\exp\biggl\{ i \pi \tau (\lambda+ \beta)_+^2 +
i \pi \bar \tau (\lambda+ \beta)_-^2
&
+ 2 \pi i (\lambda+\beta, \xi) - 2 \pi i
(\lambda+\half \beta, \alpha) \biggr\} \cr
= & e^{i \pi (\beta,\alpha)}
 \exp[{ \pi \over  2 y} ( \xi_+^2 - \xi_-^2) ] \cr
\sum_{\lambda\in \Lambda + \gamma}
\exp\biggl\{ i \pi \tau (\lambda+ \beta)_+^2 +
i \pi \bar \tau (\lambda+ \beta)_-^2
&
+ 2 \pi i (\lambda+\beta, \xi) - 2 \pi i
(\lambda+  \beta, \alpha) \biggr\} \cr}
}
where $y=\imt$.

The main transformation law is:
\eqn\thetess{
\Theta_{\Lambda } (-1/\tau, \alpha,\beta; P, {\xi_+ \over  \tau} +
{\xi_- \over  \bar \tau} )
= \sqrt{\vert \Lambda \vert \over  \vert \Lambda' \vert}
(-i \tau)^{b_+/2} (i \bar \tau)^{b_-/2}
\Theta_{\Lambda' } ( \tau, \beta,-\alpha ; P, \xi )
}
where $\Lambda'$ is the dual lattice.
If there is a characteristic vector, call it $w_2$, such that
\eqn\characteristic{
(\lambda,\lambda) = (\lambda, w_2)~ \mod ~2
}
for all $\lambda$
then we have in addition:
\eqn\thettee{
\Theta_{\Lambda } (\tau+1, \alpha,\beta; P, \xi)
= e^{-i \pi(\beta,w_2)/2}
\Theta_{\Lambda } ( \tau, \alpha-\beta-\half w_2,\beta ; P, \xi )
}

\listrefs

\bye